\documentclass[
twocolumn,  
secnumarabic,amssymb, nobibnotes, aps, prd
,superscriptaddress,
nofootinbib,
]{revtex4-1}

\usepackage[ddmmyyyy]{datetime} 
\usepackage{amsmath} 
\usepackage[dvipsnames]{xcolor}
\usepackage{cancel} 
\usepackage{graphicx}
\usepackage{comment} 
\usepackage{soul,ulem} 
\soulregister\cite7
\soulregister\ref7 
\soulregister\eqref7 
\usepackage{bbding}
\usepackage{pifont}
\usepackage{wasysym}
\usepackage{amssymb}
\usepackage{textcomp} 

\definecolor{mypurple}{RGB}{135,10,190} 

\begin{document}

\title{Jointly fitting weak lensing, X\,-\,ray and Sunyaev\,-\,Zel'dovich data\\ to constrain scalar\,-\,tensor theories with clusters of galaxies}

\author{Vincenzo F. Cardone}
\email{vincenzo.cardone@inaf.it}
\affiliation{I.N.A.F. - Osservatorio Astronomico di Roma, Via Frascati 33, 00040, Monteporzio Catone, Roma, Italy}
\affiliation{I.N.F.N. - Sezione di Roma 1, P.le Aldo Moro 2, 00185, Roma, Italy}

\author{Purnendu Karmakar}
\email{purnendu.karmakar@pd.infn.it}
\affiliation{Dipartimento di Fisica, Sapienza Universit\'a di Roma, P.le Aldo Moro 2, 00185, Roma, Italy}

\author {Marco De Petris}
\email{marco.depetris@roma1.infn.it}
\affiliation{Dipartimento di Fisica, Sapienza Universit\'a di Roma, P.le Aldo Moro 2, 00185, Roma, Italy}
\affiliation{I.N.F.N. - Sezione di Roma 1, P.le Aldo Moro 2, 00185, Roma, Italy}
\affiliation{I.N.A.F. - Osservatorio Astronomico di Roma, Via Frascati 33, 00040, Monteporzio Catone, Roma, Italy}

\author {Roberto Maoli}
\email{roberto.maoli@roma1.infn.it}
\affiliation{Dipartimento di Fisica, Sapienza Universit\'a di Roma, P.le Aldo Moro 2, 00185, Roma, Italy}



\begin{abstract}

Degenerate higher\,-\,order scalar\,-\,tensor (DHOST) theories are considered the most general class of scalar\,-\,tensor theories so that any constraints on them apply to the full set of scalar\,-\,tensor models. DHOST theories modify the laws of gravity even at galaxy clusters scale hence affecting the weak lensing (WL), X\,-\,ray and Sunyaev\,-\,Zel'dovich observables. We derive the theoretical expression for the lensing convergence $\kappa$, and the pressure profile $P$, of clusters in the framework of DHOST theories, and quantify how much they deviate from their General Relativity (GR) counterparts. We argue that combined measurements of $\kappa$, $P$, and of the electron number density, $n_e$, can constrain both the cluster and DHOST theory parameters. We carry on a Fisher matrix forecasts analysis to investigate whether this is indeed the case considering different scenarios for the spatial resolution and errors on the measured quantities.

\end{abstract}



\date{\today\ at \currenttime\ of Roma} 
\maketitle



\section{Introduction}

Cosmic acceleration may be intriguingly taken as first evidence of a failure in our understanding of gravity. Rather than relying on standard General Relativity (GR), one could then consider departure from it as it is done, e.g., in scalar\-\,tensor theories. In particular, the degenerate higher\,-\,order scalar,\-\,tensor theory (hereafter, DHOST) represents the most general scalar,\-\,tensor theory, including a propagating scalar and two tensor degrees of freedom given under the general covariance \cite{Langlois:2015cwa,Crisostomi:2016czh,Achour:2016rkg,Motohashi:2016ftl,BenAchour:2016fzp}. What makes DHOST still more attractive is that a large numbers of modified gravity (MG) models can be seen obtained as sub\,-\,cases of the most general DHOST one. This holds true for the Brans-Dicke theory \cite{Brans:1961sx}, $f(R)$ gravity \cite{DeFelice:2010aj,Nojiri:2010wj}, covariant Galileon \cite{Deffayet:2009wt,Neveu:2013mfa,Neveu:2016gxp}, Horndeski \cite{Horndeski:1974wa,Kobayashi:2011nu}, transforming gravity \cite{Zumalacarregui:2013pma}, and GLPV theory \cite{Gleyzes:2014dya}). DHOST can therefore be taken as a general framework to investigate the impact of deviations from GR on observables at different scales. 

The additional scalar degrees of freedom present in most scalar\,-\,tensor theories often give rise to an effective fifth force on the Solar system scale which must be suppressed in order to not spoil down the success of GR on such small scales. Alternatively, a screening mechanism must be invoked such as the Vainshtein screening \cite{Vainshtein:1972sx, Babichev:2013usa,Brax:2004qh} originating from the non-linear self\,-\,interaction. Such screening, however, may fail for the subset of DHOST models selecting by the condition that the propagation of gravitational waves (GWs) happens at the same speed as light \cite{Crisostomi:2017lbg,Langlois:2017dyl,Bartolo:2017ibw,Dima:2017pwp,Hirano:2019scf,Crisostomi:2019yfo}. 

In the effective field theory (EFT) description of the DHOST models, six time dependent functions $(\alpha_M, \alpha_B, \alpha_K, \alpha_T, \alpha_H, \beta_1)$ have to be assigned. The nearly contemporary arrival of the GW event GW170817 and of the light emitted by its electromagnetic counterpart GRB170817A puts the stringent constraint $|c_g^2/c^2 - 1| \lesssim 10^{-15}$ \cite{TheLIGOScientific:2017qsa, Monitor:2017mdv}, being $(c_g, c)$ the speed of GW and light, respectively. This allows to constrain $\alpha_T$ \cite{Nishizawa:2014zna,Lombriser:2015sxa,Ezquiaga:2017ekz, Sakstein:2017xjx,Creminelli:2017sry,Baker:2017hug, Jain:2015edg, Arai:2017hxj,Gumrukcuoglu:2017ijh,Oost:2018tcv,Gong:2018cgj,Gong:2018vbo}, while a further reduction of the parameter space is possible imposing stability and no\,-\,ghost conditions. Motivated by these considerations, we will henceforth concentrate on a subset of DHOST theories characterized by the two free functions $(\alpha_H, \beta_1)$ also allowing for departure of the effective gravitational constant from the Newtonian one. 

As a general comment, we note that two ingredients are needed to test a whatever gravity theory on astrophysical scales. First, we need an observable which can be predicted by the theory given a set of available auxiliary quantities. Second, the theoretical prediction must be contrasted with its measured counterpart. Galaxy clusters are ideal testing laboratories from this point of view because of the various kind of data which can be inferred from observations in different wavebands. X\,-\,ray images allow to reconstruct their gas (or IntraCluster Medium, ICM) density profile which can then be input to the hydrostatic equilibrium equation (in the form predicted by the gravity theory) to predict the pressure profile once a model for the dark halo is assumed. The pressure profile can then be inferred from Sunyaev\,-\,Zel'dovich (SZ) signal towards the clusters \cite{Sunyaev:1972eq}. Indeed, the Compton parameter is proportional to the ICM pressure along the line of sight making it possible to constrain both the cluster parameters and the theory of gravity.

Moreover, the halo density profile can be further constrained through shear measurements hence breaking degeneracy among model parameters. 

Motivated by the above qualitative sketch, we investigate here which constraints can be put on the DHOST theories parameters by jointly fitting the electron number density $n_{e}(r)$ measured from X\,-\,ray data, the pressure profile $P(r)$ inferred from SZ observations, and the lensing convergence $\kappa(R)$ reconstructed from optical data. We employ what is somewhat referred to as the {\it backward} method (see, e.g., \cite{Ettori:2018tus}), i.e., we assume parametric models for $n_{e}(r)$ and the dark halo density profile $\rho(r)$, and compute the theoretical convergence and pressure profiles. The model and DHOST parameters are then constrained by fitting the $(\kappa, n_e, P)$ data\footnote{It is worth noting that, while the convergence is directly measured on 2D maps (with $R$ the cylindrical radius), the pressure profile $P(r)$ is defined in the 3D space, but obtained by deprojecting what is measured on the 2D $y$ maps. As such, the convergence $\kappa$ is less prone to assumptions about the intrinsic symmetry properties of the system, while spherical symmetry is implicitly assumed in the deprojection needed to infer the measured pressure data.} To determine the accuracy which can be achieved by this method, we perform a Fisher matrix forecast analysis based on realistic assumptions for the signal\,-\,to\,-\,noise (S/N) of the data. We also investigate how the results depend on the adopted observational specifics varying both the sampling and the overall S/N amplitude.

The plan of the paper is as follows. Sect.\,II describes the deviations of the gravitational potentials from the GR ones due to the adoption of DHOST theories. These results are then used in Sect.\,III to compute the galaxy clusters observables of interest for our aims and to show which impact the corrections, due to the additional DHOST contributions, have on them. A theoretical ICM pressure profile is compared with the universal pressure profile, usually adopted to model SZ signal in clusters in Sect.\,IV where we also present the sample we take as input to the Fisher matrix analysis. The corresponding formalism is given in Sect.\,V, while results are discussed in Sect.\,VI before concluding in Sect.\,VII. Some supplementary material is relegated to the appendix sections.


\section{DHOST theories in the weak field limit} \label{sec:dhostforces}

The Vainshtein screening mechanism operating in DHOST theories prevents deviations from GR outside the source, but not within the source itself. On this scale, the gravitational potentials are modified. In the weak field limit, for a static spherically symmetric object, the modified Newtonian potential ($\Phi$) and curvature perturbation ($\Psi$) indeed read\cite{Langlois:2017dyl,Crisostomi:2017lbg, Bartolo:2017ibw,Dima:2017pwp} 

\begin{eqnarray}
\frac{d\Phi(r)}{dr} & = & \frac{G_{N}^{\rm eff} M(<r)}{r^2} + \Xi_1 G_{N}^{\rm eff} \frac{d^2M(<r)}{dr^2} \ , 
\label{eq: gravforceone} \\
\nonumber \\
\frac{d\Psi(r)}{dr} & = & \frac{G_{N}^{\rm eff} M(<r)}{r^2} + \frac{\Xi_2 G_{N}^{\rm eff}}{r} \frac{dM(<r)}{dr} \nonumber \\ 
 & + & \Xi_3 G_{N}^{\rm eff} \frac{d^2M(<r)}{dr^2} \ ,
\label{eq: gravforcetwo}
\end{eqnarray}
where $M(<r)$ is the total cumulative mass within the radius $r$, $G_{N}^{\rm eff}$ is the effective gravitational constant, the dimensionless coefficients
$(\Xi_1, \Xi_2, \Xi_3)$ are related to the value (at the system redshift) of the functions entering the DHOST Lagrangian. In terms of the EFT parameters, the effective gravitational constant reads

\begin{eqnarray}
G_{N}^{\rm eff} & = & [16 \pi F (1 + \Xi_0)]^{-1} \nonumber \\ 
 & = & [8 \pi (M_{Pl}^{DHOST})^2 (1 + \Xi_0)]^{-1} \nonumber \\ 
 & = & \gamma_N G_N/(1 - \alpha_H - 3 \beta_1)
\label{eq: gneff}
\end{eqnarray}
where $F$ is the coupling function between matter and geometry, and $\Xi_0 $ has been expressed as a function of $(\alpha_H, \beta_1)$ and of the DHOST Planck mass $M_{Pl}^{DHOST}$ which can be conveniently rewritten as $M_{Pl}^{DHOST} = M_{Pl}/\gamma_N$ with $\gamma_N$ as an easier to handle parameter. It is possible to relate $\gamma_N$ to the quantities setting the background expansion of the universe, while $(\alpha_H, \beta_1)$ determine the perturbative behaviour. The EFT functions also determine the amplitude of the non Newtonian terms in Eqs.(\ref{eq: gravforceone})\,-\,(\ref{eq: gravforcetwo}) being

\begin{equation}
\Xi_1 = -\frac{(\alpha_H + \beta_1)^2}{2(\alpha_H + 2\beta_1)} \ ,
\label{eq: xi1def}
\end{equation}

\begin{equation}
\Xi_2 = \alpha_H
\label{eq: xi2def}
\end{equation}

\begin{equation}
\Xi_3 = -\frac{\beta_1(\alpha_H + \beta_1)}{2(\alpha_H + 2\beta_1)} \ .
\label{eq: xi3def}
\end{equation}
Deviations from GR can therefore be given in terms of the three parameters $(\alpha_H, \beta_1, \gamma_N)$. In the GR limit, they take the values $(1, 0, 0)$ so that Eqs.(\ref{eq: gravforceone})\,-\,(\ref{eq: gravforcetwo}) go back to the well known classical expressions and Newton gravity is restored.

As an important remark, we stress that $(\alpha_H, \beta_1, \gamma_N)$ are not constants, but actually functions of the redshift. As such, the impact of the DHOST modifications to the force law is not the same for objects at different redshifts. Moreover, since most DHOST theories reduce to GR at high $z$ in order to be in agreement with Cosmic Microwave Background (CMB) data, one expects that the larger is $z$, it will be challenging to detect deviations from GR force laws in distant clusters. This has remarkable consequences for the analysis we are interested in here. Since the DHOST correction is redshift dependent, one should be careful when stacking clusters based on $z$. Indeed, in doing that, one is assuming that the variation of the parameters $(\alpha_H, \beta_1, \gamma_N)$ is negligible over the redfshift range probed by the clusters one is willing to stack. Whether such an assumption is valid or not depends on the width of the redshift bin and the particular class of DHOST theories under investigation which is a further point to be taken into account when comparing to observations or inferring limits on the theory itself.


\section{Galaxy clusters observables}

The modified force laws impact any observable on galaxy cluster scales. However, the effect will be different depending on which of the two potentials $(\Phi, \Psi)$ enters the games. Indeed, this offers an intriguing opportunity to break some degeneracy among astrophysical and DHOST parameters so that in the following we will derive the modified expressions for quantities which can be measured from cluster observations in different frequency bands.

\subsection{Mass profile}

As a preliminary step, it is worth discussing which model we are going to use for describing the cluster cumulative mass profile since, as Eqs.(\ref{eq: gravforceone})\,-\,(\ref{eq: gravforcetwo}) show, it plays a key role. Ideally, we should include ICM, stars, and dark matter. However, the stellar mass fraction is no larger than $10\%$ in the inner kpc to then degrade quickly, while the gas contribution, although larger than the stars one, is still subdominant. We can therefore identify the total mass with the dark halo one hence modeling this component only since it accounts for more than $\sim 85\%$ of the total mass. Following the standard approach, we model it with the NFW profile whose density law is \cite{Navarro:1995iw,Navarro:1996gj}

\begin{equation}
\rho(r) = \frac{\rho_s}{(r/r_s) (1 + r/r_s)^2}
\label{eq: nfwrho}
\end{equation}
with $\rho_s$ a characteristic density, and $r_s$ the radius where the logarithmic slope $s = d\ln{\rho}/d\ln{r}$ takes the isothermal value $s = -2$. Under the assumption of spherical symmetry, the mass profile can be straightforwardly obtained and it is conveniently rewritten as 

\begin{equation}
M(<r) = M_{\Delta} \frac{\ln{(1 + c_\Delta x)} - c_\Delta x/(1 + c_\Delta x)}{\ln{(1 + c_\Delta)} - c_\Delta/(1 + c_\Delta)}
\label{eq: massnfw}
\end{equation}
with $x = r/R_\Delta$, $R_\Delta$ the radius which the mean mass density is $\Delta$ times the universe critical one $\rho_c(z)$ at the halo redshift $z$, $c_{\Delta} = R_{\Delta}/r_s$ the halo concentration, and 

\begin{equation}
M_\Delta = \frac{4}{3} \pi \Delta \rho_{c}(z) R_{\Delta}^{3} \ .
\label{eq: mdeltadef}
\end{equation}
Following common approach in the literature, we replace $(\rho_s, r_s)$ with $(M_{\Delta}, c_{\Delta})$ as model parameters. Different choices are possible for $\Delta$. In particular, unless otherwise stated, we will set $\Delta = 200$, and refer to $M_{\Delta}$ as the halo mass although formally the halo may expand beyond $R_{200}$. In X\,-\,ray and SZ studies, the choice $\Delta = 500$ is preferred since typically data cover up to $1 - 2 R_{500}$ (or $3 R_{500}$ relying on Planck measurements) so that one can estimate $M_{500}$ rather than $M_{200}$. Similarly, one can define a concentration $c_{500} = R_{500}/r_s$ which can be found once the model parameter $c_{200}$ is set by solving\footnote{Hereafter, we will use $\ln{x} (\log{x})$ to denote the natural (base 10) logarithm.}  

\begin{equation}
\frac{5}{2} \left ( \frac{c_{500}}{c_{200}} \right )^3 = 
\frac{\ln{(1 + c_{500})} - c_{500}/(1 + c_{500})}
{\ln{(1 + c_{200})} - c_{200}/(1 + c_{200})} \ .
\label{eq: c500vsc200}
\end{equation}
For later applications, it is convenient to write down the first three derivatives of the mass profile. It is only a matter of algebra to get

\begin{eqnarray}
\frac{dM(<r)}{dr} & = & \frac{M_{200}}{R_{200}} \frac{c_{200}^2}{\ln{(1 + c_{200})} - c_{200}/(1 + c_{200})} \nonumber \\ 
 & \times & \frac{x}{(1 + c_{200} x)^2} \ ,
\label{eq: dmdrnfw}
\end{eqnarray}

\begin{eqnarray}
\frac{d^2M(<r)}{dr^2} & = & \frac{M_{200}}{R_{200}^2} \frac{c_{200}^2}{\ln{(1 + c_{200})} - c_{200}/(1 + c_{200})} \nonumber \\ 
 & \times & \frac{1 - c_{200} x}{(1 + c_{200} x)^3} \ ,
\label{eq: d2mdr2nfw}
\end{eqnarray}

\begin{eqnarray}
\frac{d^3M(<r)}{dr^3} & = & \frac{M_{200}}{R_{200}^3} \frac{c_{200}^2}{\ln{(1 + c_{200})} - c_{200}/(1 + c_{200})} 
\nonumber \\ 
 & \times & \frac{2 c_{200} (c_{200} x - 2)}{(1 + c_{200} x)^4} \ .
\label{eq: d3mdr3nfw}
\end{eqnarray}
It is worth noting that, because of Eq.(\ref{eq: mdeltadef}), $M_{200}$ cancels out from the ratio $M_{200}/R_{200}^3$ so that one could naively infer that the third derivative of the mass profile could be independent on the halo mass $M_{200}$. This is actually not the case since the residual dependence on $M_{200}$ is hidden into $x = r/R_{200}$. Moreover, as we will see, one can postulate that the mass\,-\,concentration relation $c_{200} = c_{200}(M_{200}, z)$ which is found in GR based N\,-\,body simulations also holds in DHOST theories hence introducing a further dependence on $M_{200}$.

\subsection{Lensing convergence profile}

Gravitational lensing is able to probe the mass distribution along the line of sight. The deflection angle can be expressed as the derivative of the effective lensing potential integrated along the line of sight. For any metric theory, this is given by

\begin{equation}
\Phi_{lens}(R) = \frac{2}{c^2} \frac{D_{LS}}{D_L D_S} \int_{-\infty}^{+\infty}{\frac{\Phi(R, \ell) + \Psi(R, \ell)}{2} d\ell}
\label{eq: lenspot}
\end{equation}
where $(R, \ell)$ are the usual cylindrical coordinates with $\ell$ the axis along the line of sight. In Eq.(\ref{eq: lenspot}), $(D_L, D_S, D_{LS})$ are the angular diameter distances from the observer to the lens, the observer to the source, and between lens and source, respectively, fixing the geometry of the system. For a spherically symmetric lens, the convergence profile can then be computed as

\begin{equation}
\kappa(R) = \frac{1}{c^2} \frac{D_{LS}}{D_L D_S} \int_{-\infty}^{+\infty}{\nabla_r \left [\frac{\Phi(R, \ell) + \Psi(R, \ell)}{2} \right ] d\ell}
\label{eq: kappadef}
\end{equation}
where $\nabla_r = \partial^2/\partial r^2 + (2/r) \partial/\partial_r$ is the Laplacian with respect to $r = (R^2 + \ell^2)^{1/2}$. Using the general Eqs.(\ref{eq: gravforceone})\,-\,(\ref{eq: gravforcetwo}), we get

\begin{eqnarray}
\nabla_r \Phi(r) & = & \frac{G_{N}^{\rm eff}}{r^2} \frac{dM(<r)}{dr} 
\nonumber \\ & + & 
\frac{2 \Xi_1 G_{N}^{\rm eff}}{r} \frac{d^2M(<r)}{dr^2} 
+ \Xi_1 G_{N}^{\rm eff} \frac{d^3M(<r)}{dr^3}
\nonumber \\
\nabla_r \Psi(r) & = & \frac{(1 + \Xi_2) G_{N}^{\rm eff}}{r^2} \frac{dM(<r)}{dr} \nonumber \\ & + & \frac{(\Xi_2 + 2 \Xi_3) G_{N}^{\rm eff}}{r} \frac{d^2M(<r)}{dr^2} 
\nonumber \\ & + & 
\Xi_3 G_{N}^{\rm eff} \frac{d^3M(<r)}{dr^3} \nonumber
\label{eq: gravforceder}
\end{eqnarray}
so that, using Eqs.(\ref{eq: massnfw})\,-\,(\ref{eq: d3mdr3nfw}) for the NFW profile, we get

\begin{eqnarray}
\nabla_r \Phi_{lens} & = & 
\frac{\gamma_N G_{N} M_{200}}{R_{200}^3} \nonumber \\ 
 & \times & \frac{c_{200}^2}{\ln{(1 + c_{200})} - c_{200}/(1 + c_{200})} \frac{1}{x (1 + c_{200} x)^2} \nonumber \\ \nonumber \\ 
 & \times & 
\left \{ 1 + \frac{\alpha_H}{1 + c_{200} x} - \frac{\alpha_H + \beta_1}{4} \frac{2 - 4 c_{200} x}{(1 + c_{200} x)^2} \right \} \ \ 
\label{eq: nablaphilens}
\end{eqnarray}
where it is 

\begin{equation}
\Phi_{lens}(R, z) = \frac{\Phi(R, z) + \Psi(R, z)}{2} \ ,
\label{eq: philens}
\end{equation}
and we remind $x = r/R_{200}$, and we have explicitly introduced the DHOST parameters $(\alpha_H, \beta_1, \gamma_N)$. Inserting Eq.(\ref{eq: nablaphilens}) into Eq.(\ref{eq: kappadef}) and changing integration variable from $\ell$ to $\zeta = \ell/R_v$, we finally get the following expression for the DHOST convergence 

\begin{equation}
\kappa(R) = \frac{\gamma_N \Sigma_{200}}{\Sigma_{crit}}
\frac{c_{200}^{2} {\cal{K}}_0(\xi, {\bf p})}
{\ln{(1 + c_{200})} - c_{200}/(1 + c_{200})}
\label{eq: kappadhost}
\end{equation}
with $\xi = R/R_{200}$, ${\bf p} = \{c_{200}, \alpha_H, \beta_1\}$, $\Sigma_{200} = M_{200}/4 \pi R_{200}^2$, $\Sigma_{crit} = c^2 D_S/(4 \pi G_N D_L D_{LS})$ the critical surface density for lensing, and

\begin{equation}
{\cal{K}}_0(\xi, {\bf p}) = 
\int_{-\infty}^{\infty}{\frac{{\cal{S}}(\xi, \zeta, {\bf p})}{(\xi^2 + \zeta^2)^{1/2} [1 + c_{200} (\xi^2 + \zeta^2)^{1/2}]^2} d\zeta}
\label{eq: kappatildedef}
\end{equation}
having defined 

\begin{eqnarray}
{\cal{S}}(\xi, \zeta, {\bf p}) & =  & 1 + \frac{\alpha_H}{1 + c_{200} (\xi^2 + \zeta^2)^{1/2}} \nonumber \\
& - & \frac{\alpha_H + \beta_1}{4} \frac{2 - 4 c_{200} (\xi^2 + \zeta^2)^{1/2}}{[1 + c_{200} (\xi^2 + \zeta^2)^{1/2}]^2}  \ .
\label{eq: calsdef}
\end{eqnarray}
Some comments are in order here. First, as a consistency check, we note that setting $(\alpha_H, \beta_1, \gamma_N) = (0, 0, 1)$ gives back the GR result as expected. Second, we note that, in the very inner regions (i.e., for $r << r_s$ hence $c_{200} (\xi^2 + \zeta^2)^{1/2} << 1$), it is 

\begin{displaymath}
{\cal{S}}(\xi, \zeta) \sim 1 - (\alpha_H - \beta_1)/2 
\end{displaymath}
so that the net effect is to rescale the virial mass from $M_{200}$ to $M_{200} [ 1 - (\alpha_H - \beta_1)/2]$. In the opposite asymptotic limit $r >> r_s$ (i.e., $c_{200} (\xi^2 + \zeta^2)^{1/2} >> 1$), it is on the contrary 

\begin{displaymath}
{\cal{S}}(\xi, \zeta) \sim 1 + (2 \alpha_H - \beta_1)/(\xi^2 + \zeta^2)^{1/2}
\end{displaymath}
showing that the deviation from GR quickly fades away hence making it harder to detect it. We therefore expect that the convergence profile data will be more sensible to the DHOST parameters in the intermediate region (i.e., $R \sim r_s)$ where it is possible to both appreciate the correction and break the degeneracy between $\alpha_H$ and $\beta_1$ thanks to the different scaling of the second and third term of ${\cal{S}}(\xi, \zeta)$ in this regime. On the contrary, the $\gamma_N$ parameter only appears in the product $\gamma_N M_{200}/R_{200}^2 \propto \gamma_N M_{200}^{1/3}$ so that its effect could be absorbed by a naive rescaling of the halo mass. However, the degeneracy is partially broken by the fact that $M_{200}$ also indirectly enters through the dependence of the concentration on mass. 

\subsection{Theoretical pressure profile}

The other observable we will use as a constraint on the DHOST and cluster parameters comes from observations of the SZ signal which allows to recover the pressure profile. As hinted at in the Introduction, its theoretical counterpart can be derived once models are assumed for the halo mass profile and the electron number density. Indeed, under the assumption of hydrostatic equilibrium, one has

\begin{equation}
\frac{1}{\rho_{gas}(r)} \frac{dP(r)}{dr} = -\frac{d\Phi(r)}{dr}
\label{eq: dpdrhydro}
\end{equation}
where the gas density profile can be conveniently related to the electron number density $n_e(r)$ as $\rho_{gas}(r) \simeq 1.8 \mu m_p n_e(r) = m_{eff} n_e(r)$ with $m_p$ the proton mass and $\mu$ the mean molecular weight. In order to be consistent with the cluster catalog we will introduce later, we adopt the double\,-\,$\beta$ model \cite{Xue:2000uu} to set

\begin{equation}
\frac{n_{e}^{2}(r)}{n_{01}^{2}} = \left [ 1 + \left ( \frac{r}{r_{c1}} \right )^2 \right ]^{-3 \beta} + \left ( \frac{n_{02}}{n_{01}} \right )^2
\left [ 1 + \left ( \frac{r}{r_{c2}} \right )^2 \right ]^{-3\beta} 
\label{eq: betamodel}
\end{equation}
with $(\beta, r_{c1}, r_{c2}, n_{01}, n_{02})$ parameters to be inferred from the fit of X\,-\,ray data. Using Eq.(\ref{eq: gravforceone}) for $d\Phi(r)/dr$ and the NFW model for the mass profile (under the assumption that the dark halo is the main contributor to the total mass), one can integrate Eq.(\ref{eq: dpdrhydro}) with the boundary condition that the pressure vanishes at infinity to get

\begin{eqnarray}
P(x) & = & \frac{m_{eff} n_{01} \gamma_N}{1 - \alpha_H - 3 \beta_1} 
\frac{G_N M_{200}}{R_{200}} \\ 
 & \times &
\left [ {\cal{Q}}_{0}^{GR}(x, c_{200})  
- \frac{(\alpha_H + \beta_1)^2}{2 (\alpha_H + 2 \beta_1)} {\cal{Q}}_{0}^{MG}(x, c_{200}) \right ] \nonumber
\label{eq: prdhost}
\end{eqnarray}
where we have defined

\begin{equation}
\begin{array}{l}
\displaystyle{{\cal{Q}}_{0}^{GR}(x, c_{200}) =} \\
 \\
\displaystyle{\int_{x}^{\infty}{
\frac{\ln{(1 + c_{200} x^{\prime})} - c_{200} x^{\prime}/(1 + c_{200} x^{\prime})}{x^{\prime 2} [\ln{(1 + c_{200})} - c_{200}/(1 + c_{200})]}
\frac{n_{e}(x^{\prime})}{n_{01}} dx^{\prime}}}
\end{array}
\label{eq: qzgrdef}
\end{equation}

\begin{equation}
\begin{array}{l}
\displaystyle{{\cal{Q}}_{0}^{MG}(x, c_{200}) =} \\
 \\ 
\displaystyle{\int_{x}^{\infty}{
\frac{c_{200}^2 (1 - c_{200} x^{\prime}) (1 + c_{200} x^{\prime})^{-3}}
{ \ln{(1 + c_{200})} - c_{200}/(1 + c_{200})}
\frac{n_{e}(x^{\prime})}{n_{01}} dx^{\prime}}}
\end{array}
\label{eq: qzmgdef}
\end{equation}
where we have expressed the electron number density in Eq.(\ref{eq: betamodel}) in terms of the dimensionless variable $x$ taking as parameters $(x_{c1}, x_{c2}) = (r_{c1}, r_{c2})/R_{200}$. Note that the two integral functions ${\cal{Q}}_{0}^{GR}$ and ${\cal{Q}}_{0}^{MG}$ depend on the gas density parameters too, but we have not explicitly indicated them as argument just to shorten the notation.

It is worth stressing that the DHOST term in Eqs.(\ref{eq: prdhost})\,-\,(\ref{eq: qzmgdef}) depends on a combination of the $(\alpha_H, \beta_1)$ parameters other than the corresponding ones in the WL related function ${\cal{S}}(\xi, \zeta)$. Similarly, $\gamma_N$ now enters through the product $\gamma_N M_{200}/R_v \propto \gamma_N M_{200}^{2/3}$ which is again different from what takes place with the convergence profile. As a consequence, it is no more possible to rescale the halo mass to compensate for a change in $\gamma_N$ since this would fix one observable at the cost of missing the other. These considerations make us argue that a joint use of the $\kappa(R)$ and $P(r)$ profiles can break the degeneracy among DHOST parameters hence better constraining $(\alpha_H, \beta_1, \gamma_N)$. 

A final consideration about the pressure profile in Eq.(\ref{eq: prdhost}) is in order here. To get it, we have adopted the double\,-\,$\beta$ model for the electron number density profile $n_e(r)$. One could argue that all the previous works using this expression have been performed under the GR assumption so that we are extrapolating its validity outside the framework where it has been tested. However, one could consider the double\,-\,$\beta$ model simply as an empirical fitting function whose parameters must be determined by matching observations in a given framework. As such, there is no systematics induced by its adoption. Moreover, in this exploratory work, we are only interested in presenting the general formalism and apply it under realistic conditions which are guaranteed by the use of the double\,-\,$\beta$ model for $n_e(r)$. Nothing prevents to use a different profile (such as, e.g., the Vikhlinin one \cite{Vikhlinin:2005mp}) in future studies which will deal with actual data. 

\subsection{Impact of DHOST deviations from GR}

It is instructive to look at how much the modified convergence and pressure profile deviate from their GR counterparts. To this end, we must first set the halo and electron density parameters which we do by choosing three representative clusters from the sample we will introduce later in Sec.\,IV. In particular, we select MACSJ0429, MS1054, and MACSJ1423 since they are the objects with the median values of the redshift, mass, and concentration, respectively. 

\begin{figure*}
\centering
\includegraphics[width=0.65\columnwidth]{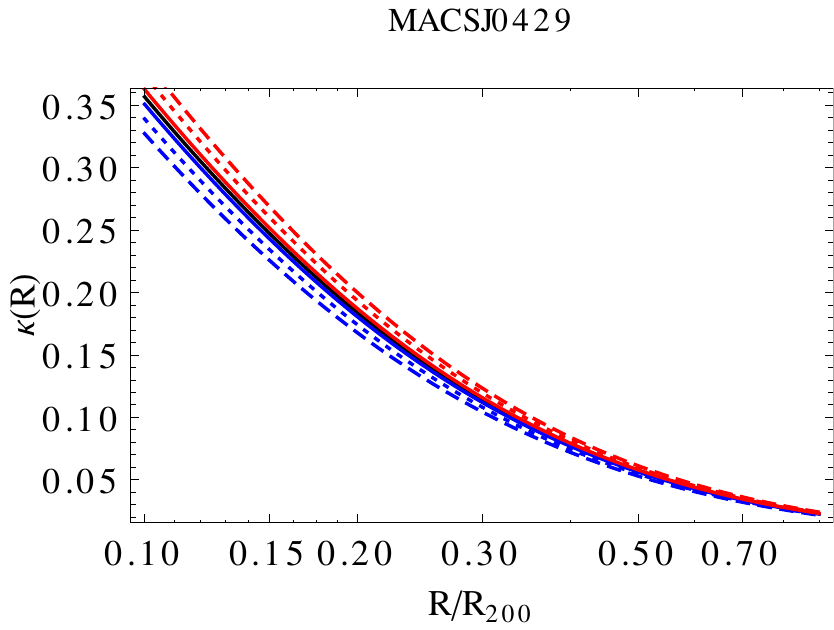}
\includegraphics[width=0.65\columnwidth]{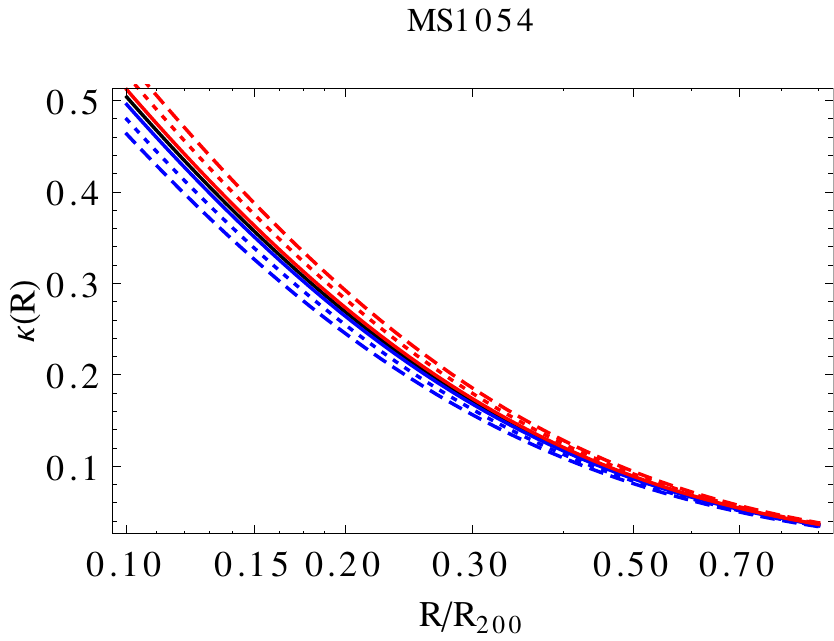}
\includegraphics[width=0.65\columnwidth]{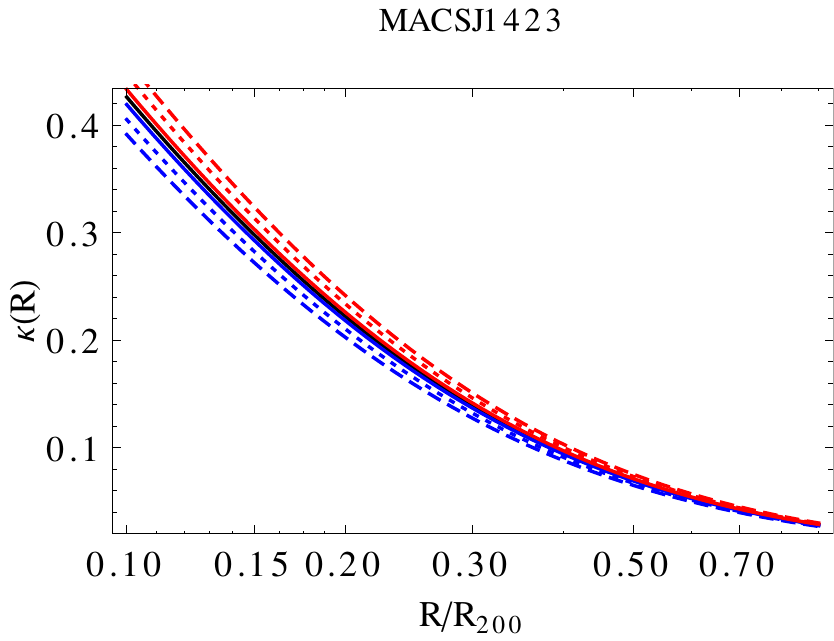} \\
\includegraphics[width=0.65\columnwidth]{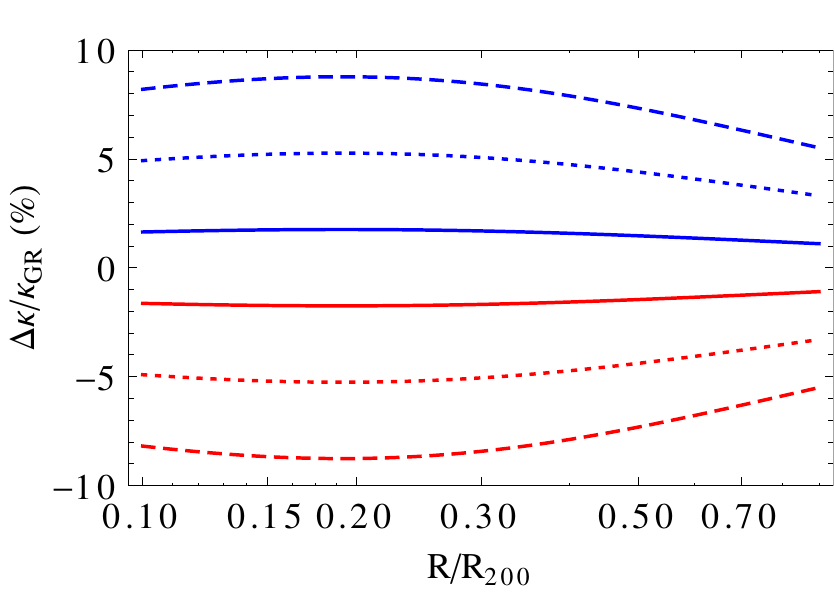}
\includegraphics[width=0.65\columnwidth]{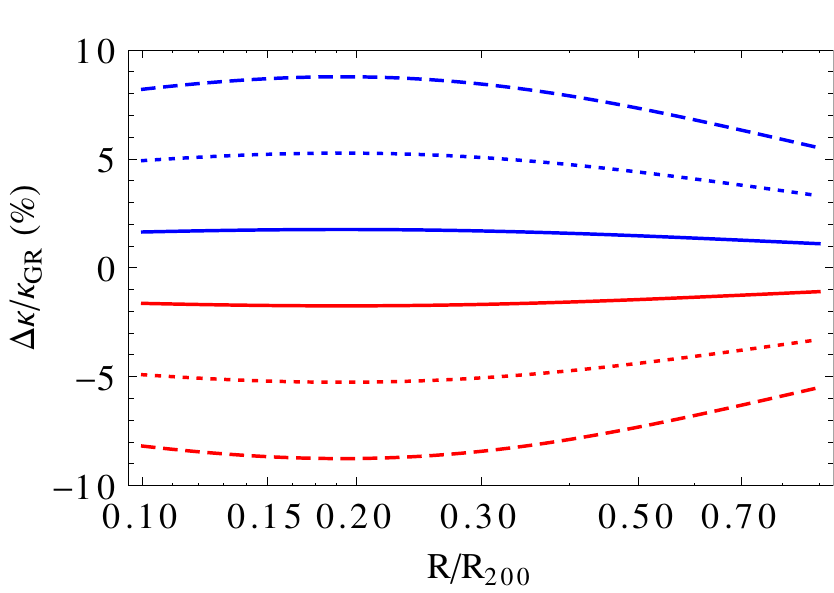}
\includegraphics[width=0.65\columnwidth]{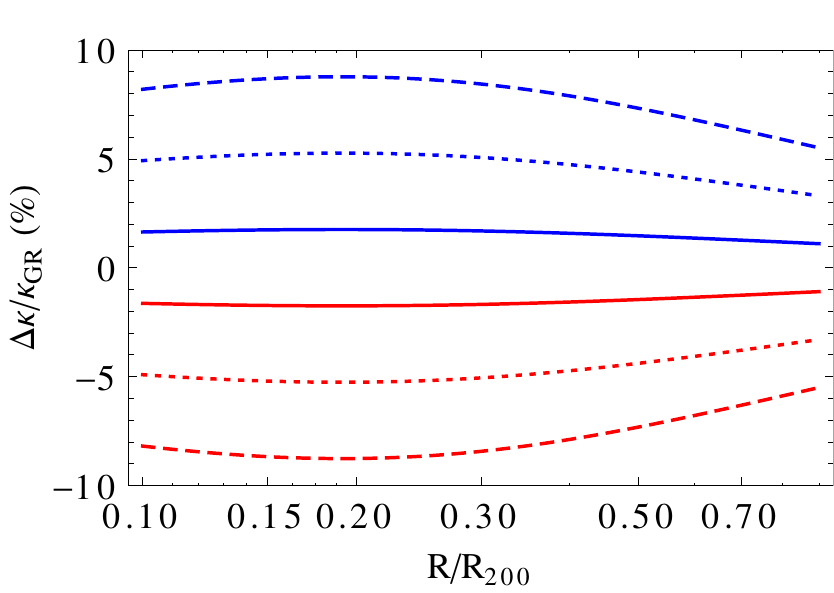} \\
\caption{{\it Top.} Lensing convergence profile for models with different values of $\alpha_H$, namely $\alpha_H = (-0.25, -0.15, -0.05)$ for blue (dashed, dotted, solid) lines, and $\alpha_H = (0.05, 0.15, 0.25)$ for red (solid, dotted, dashed) lines. For all cases, it is $\beta_1 = 0$ and $\gamma_N = 1 - \alpha_H - 3 \beta_1$. Black line is for GR. Left, centre, and right panels refer to three representative clusters. {\it Bottom.} Relative difference $\Delta \kappa/\kappa_{GR} = 100 \times (\kappa_{GR} - \kappa)/\kappa_{GR}$ for the same cases as in the top panels.}
\label{fig: kappavsalphah}
\end{figure*}

Since $\gamma_N$ only scales up or down both the convergence and the pressure theoretical profiles, understanding its impact is actually trivial so that we prefer to just focus on $\alpha_H$ and $\beta_1$. We therefore set $\gamma_N = 1 - \alpha_H - 3\beta_1$ so that the effective gravitational constant equals the Newtonian one. We will then first look at the impact of $\alpha_H$ and $\beta_1$ separately by setting one of them to zero, and varying the other over a fiducial range. This range is fixed by asking that the corresponding DHOST theory fulfils some stability criteria and its background expansion is in good accordance with the $\Lambda$CDM one. In this way we ensure that any deviation from the GR convergence and pressure is due to realistic DHOST models.

\subsubsection{Lensing convergence}

Let us first consider the convergence profile assuming the source is at $z_s = 2.0$. In Fig.\ref{fig: kappavsalphah} we plot $\kappa(R)$ for different $\alpha_H$ values, setting $\beta_1 = 0$. The sign of the difference with respect to the GR convergence can be qualitatively understood rewriting the function ${\cal{S}}$ of Eq.(\ref{eq: calsdef}) as follows

\begin{equation}
{\cal{S}}(x, \alpha_H, \beta_1 = 0) = 1 + \frac{\alpha_H}{1 + c_{200} x} \left [1 - \frac{2(1 - c_{200} x)}{1 + c_{200} x} \right ]
\end{equation}
where we have already set $\beta_1 = 0$. We thus find ${\cal{S}} > 1$ for $x > x_{min} = 1/(3 c_{200})$ if $\alpha_H > 0$ and vice versa. Noting that ${\cal{S}} > 1$ leads to an increase of the argument of the integral giving the convergence $\kappa(R)$ and considering that, for typical $c_{200}$ values, $x_{min}$ lies in the inner cluster regions, we therefore get that the DHOST convergence is larger (smaller) than the GR one (so that $\Delta \kappa/\kappa_{GR}$ is negative/positive) for $\alpha_H > 0$ ($< 0$) consistent with the results in Fig.\,\ref{fig: kappavsalphah}. The deviation fades to zero as $x$ increases because of the multiplicative term $(1 + c_{200} x)^{-1}$ with the rate of the decrease depending on the cluster concentration. Note that the dependence on $c_{200}$ is less evident in these plots since the three clusters we have chosen have quite similar concentration, being $c_{200} = (3.87, 3.38, 3.70)$ for MACSJ0429, MS1054, and MACSJ1423, respectively.

\begin{figure*}
\centering
\includegraphics[width=0.65\columnwidth]{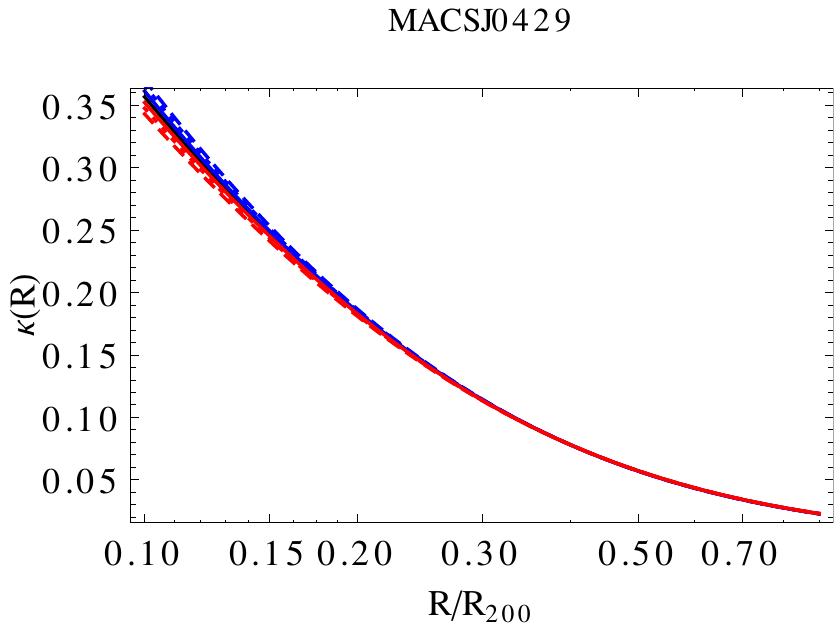}
\includegraphics[width=0.65\columnwidth]{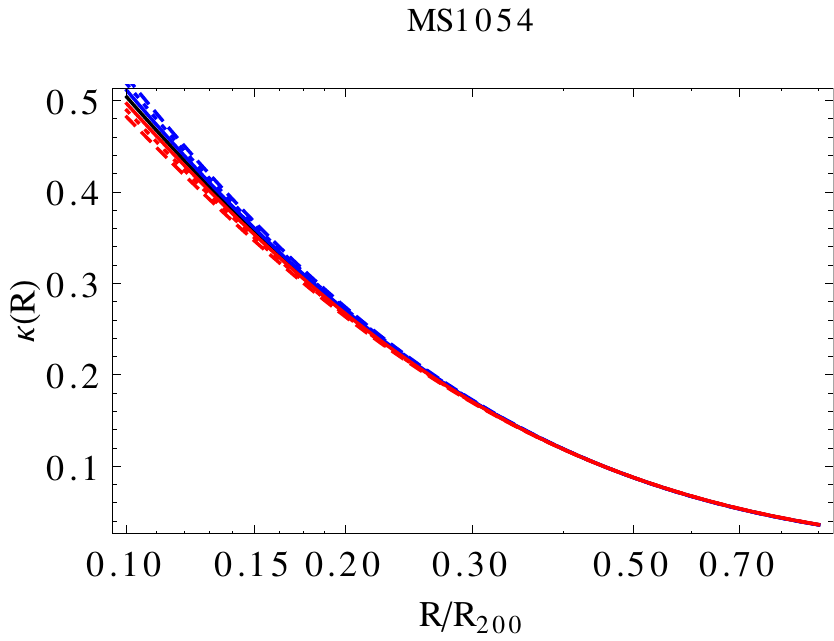}
\includegraphics[width=0.65\columnwidth]{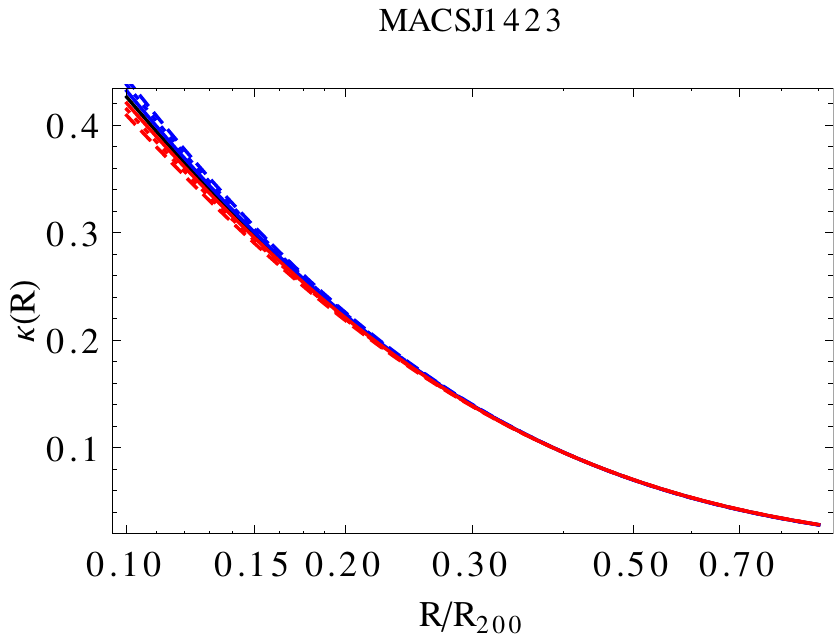} \\
\includegraphics[width=0.65\columnwidth]{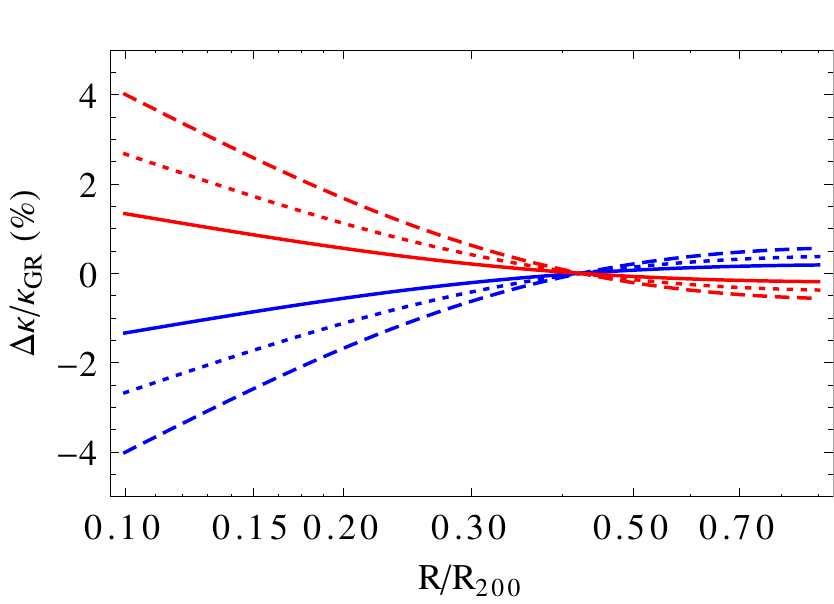}
\includegraphics[width=0.65\columnwidth]{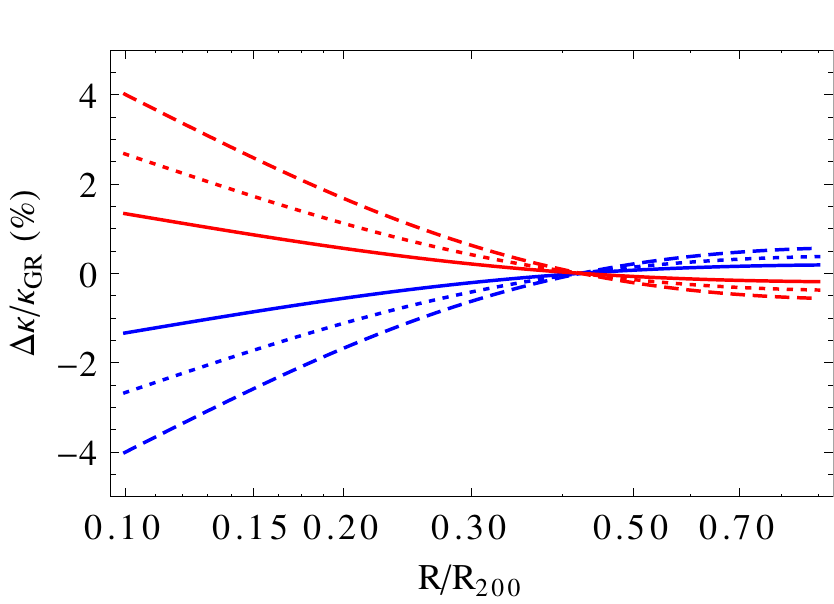}
\includegraphics[width=0.65\columnwidth]{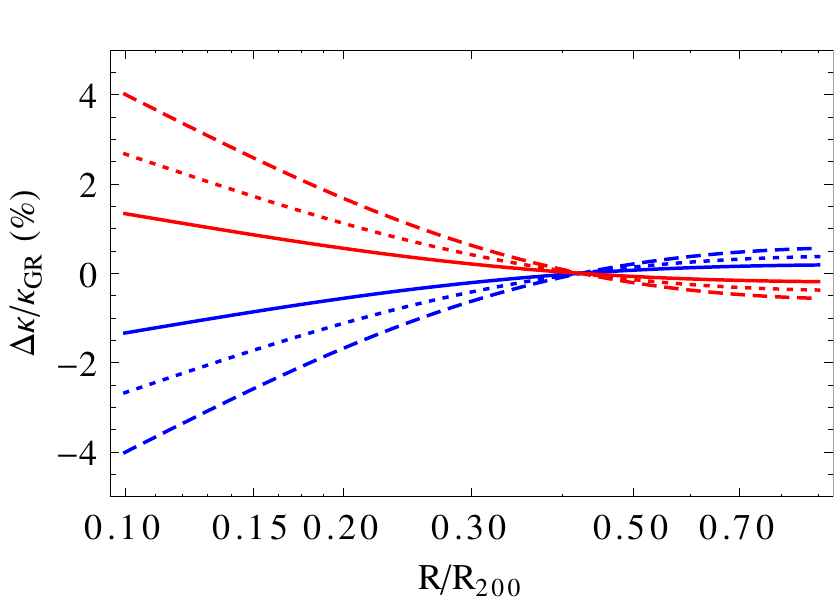} \caption{Same as Fig.\,\ref{fig: kappavsalphah} but setting $\alpha_H = 0$ and using different $\beta_1$ values, namely $\beta_1 = (-0.15, -0.10, -0.05)$ for blue (dashed, dotted, solid) lines, and $\beta_1 = (0.05, 0.10, 0.15)$ for red (solid, dotted, dashed) lines.}
\label{fig: kappavsbeta1}
\end{figure*}

\begin{figure*}
\centering
\includegraphics[width=0.65\columnwidth]{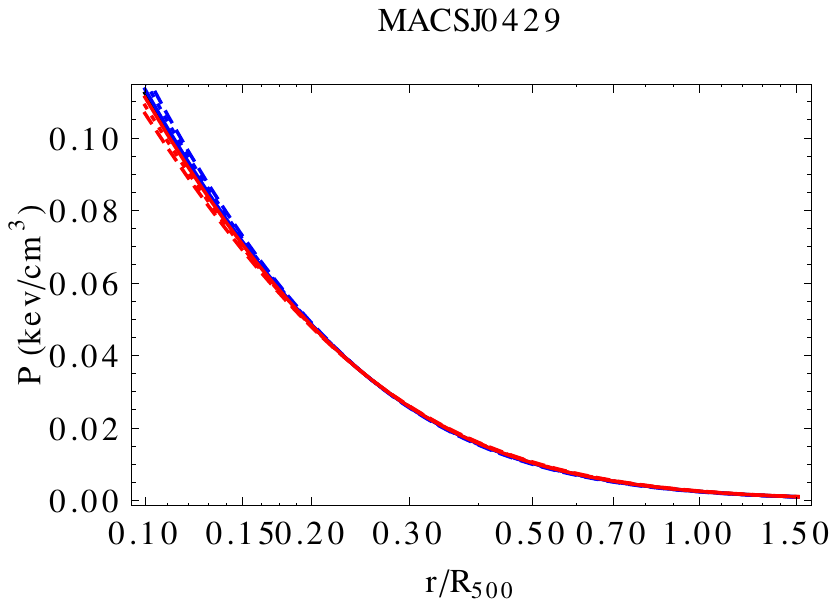}
\includegraphics[width=0.65\columnwidth]{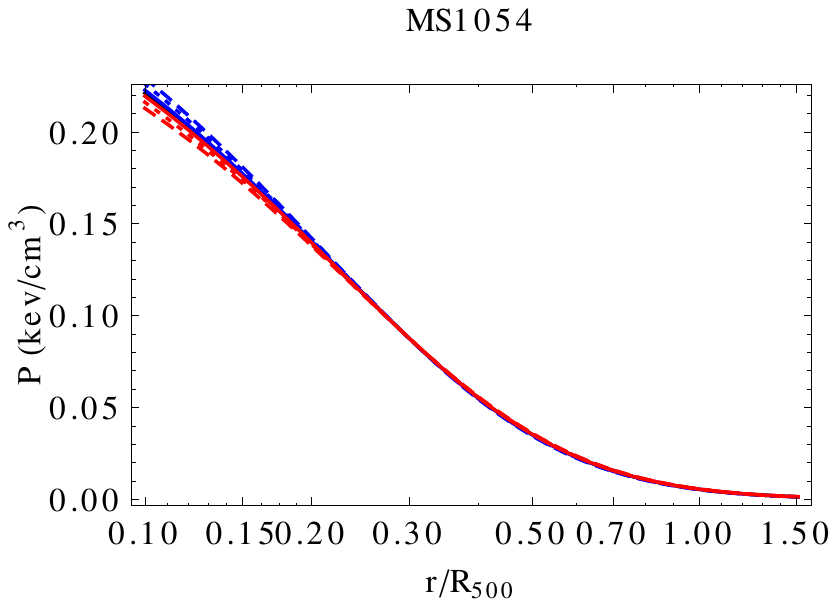}
\includegraphics[width=0.65\columnwidth]{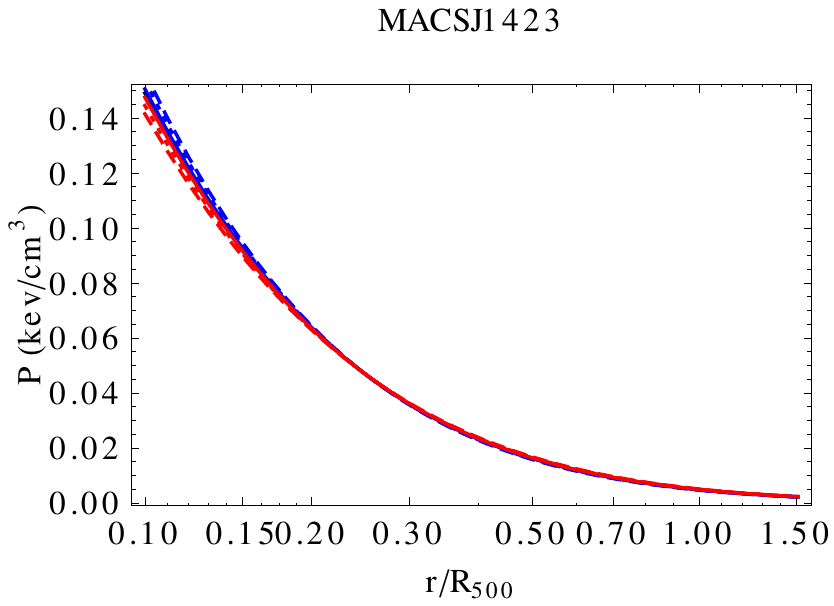} \\
\includegraphics[width=0.65\columnwidth]{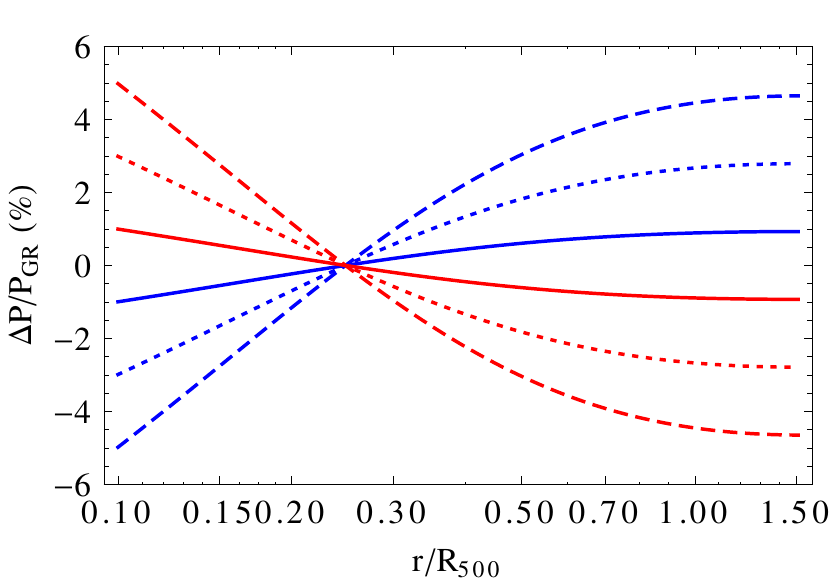}
\includegraphics[width=0.65\columnwidth]{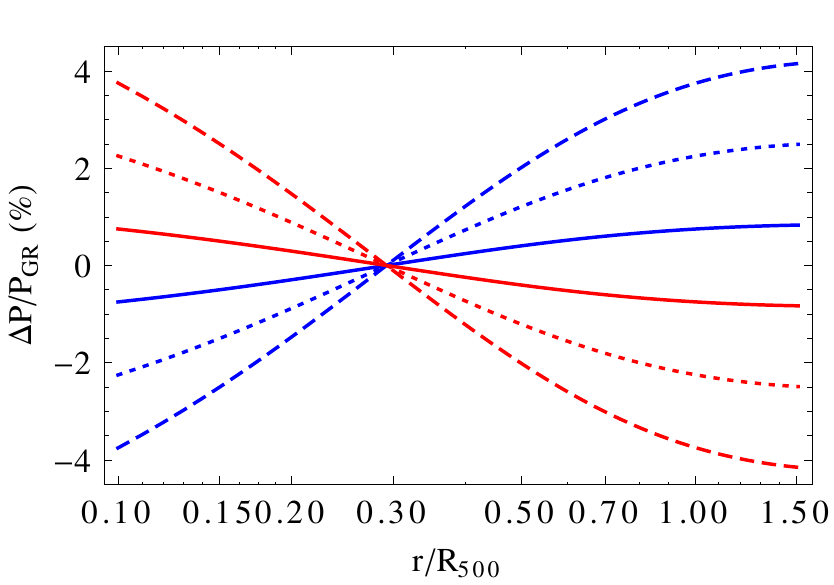}
\includegraphics[width=0.65\columnwidth]{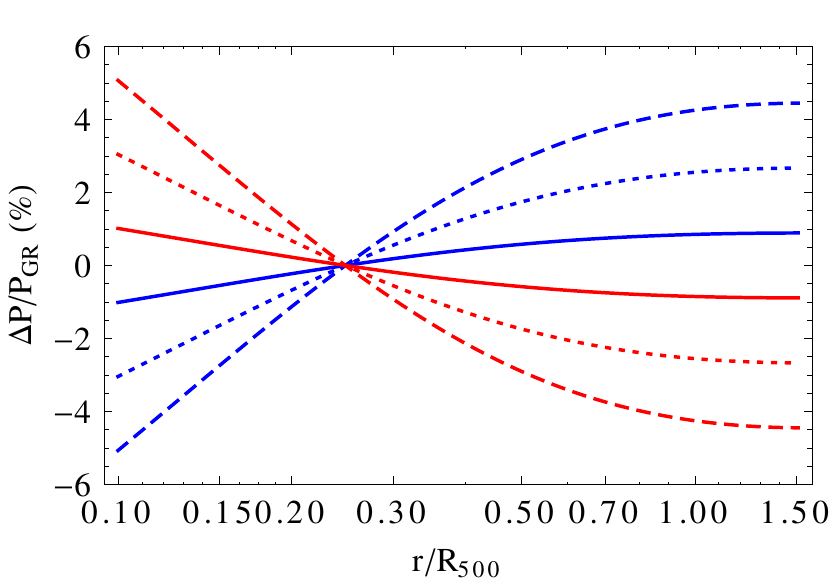} \\
\caption{Same as Fig.\,\ref{fig: kappavsalphah} but for the pressure profile $P(r)$ in top panels, and for its deviation from the GR case in the bottom panels. We set the DHOST parameters to the same values in Fig.\,\ref{fig: kappavsalphah} in order to get a fair comparison. Note that we scale the radius with respect to $R_{500} \sim R_{200}/2$ to show the radial range typically probed by SZ observations.}
\label{fig: prvsab}
\end{figure*}

Fig.\,\ref{fig: kappavsbeta1} highlights the impact of $\beta_1$ when we set $\alpha_H = 0$. In this case, we simply get

\begin{equation}
{\cal{S}}(x, \alpha_H = 0, \beta_1) = 1 - 
\frac{\beta_1 (1 - 2 c_{200} x)}{2(1 + c_{200} x)^2}
\end{equation}
so that ${\cal{S}}(x) - 1$ changes its sign at $x_{t} = 1/(2 c_{200})$ from positive to negative or vice versa depending on $\beta_1$ being positive or negative. As a consequence, $\Delta \kappa/\kappa_{GR}$ has no more a monotonic behaviour which explains the profile of the curves in Fig.\,\ref{fig: kappavsbeta1}. In Figs.\,\ref{fig: kappavsalphah} and \ref{fig: kappavsbeta1}, it is evident that $\beta_1$ has a smaller impact on the differences between GR and DHOST lensing convergence profiles. However, one should also take into account the different range allowed for the two DHOST parameters with $\beta_1$ spanning a smaller one. On the contrary, Eq.(\ref{eq: calsdef}) shows that $\beta_1$ enters the second non GR term only, while $\alpha_H$ contributes two terms with opposite sign. As a result, a change in $\beta_1$ immediately affects the convergence, while a variation of $\alpha$ is less evident because the two terms partially compensate each other. 

When we allow for both $\alpha_H$ and $\beta_1$ to change, the resulting $\Delta \kappa$ profile is qualitatively similar to the case with $\beta_1 = 0$. This is a consequence of the first additional term in Eq.(\ref{eq: calsdef}) which typically dominates over the second one where $\beta_1$ enters. As a general comment, we therefore conclude that the DHOST convergence $\kappa(R)$ may deviate from its GR counterpart by order $5 - 10\%$ over the range of radii probed by observational data. This hints at this observable as a promising probe to constrain the DHOST parameters $(\alpha_H, \beta_1)$. 

\subsubsection{Pressure profile}

Let us now consider the pressure profile $P(r)$ for the same three representative clusters considered above. Eq.(\ref{eq: qzmgdef}) shows that $(\alpha_H, \beta_1)$ only enter through the combination $\Xi_1$ as defined in Eq.(\ref{eq: xi1def}) so that a clear degeneracy among the two parameters exist. Fig.\,\ref{fig: prvsab} shows the pressure profiles and the deviations from the GR case for different values of $\alpha_H$ setting $\beta_1 = 0$. Note that we scale the distance from the centre with respect to $R_{500}$ instead of $R_{200}$ (with $R_{500} \sim 0.6 R_{200}$) in order to look at the relevant quantities over the range typically probed by actual SZ observations. 

The DHOST pressure profile turns out to be larger or smaller than the GR one depending on the sign of $\alpha_H$ being positive or negative. Both the profile and the amplitude of the deviation from GR are quite similar to those for $\Delta \kappa/\kappa_{GR}$ for $\beta_1 = 0$ although over a different radial range. This is not surprising given that, in this setting, the corrective term to the pressure has the same shape as the one for the convergence being both proportional to $(1 - c_{200} x)$. Actually, this result is not limited to the case $\beta_1 = 0$, but rather to all those cases giving the same $\Xi_1$ as the one used to get Fig.\ref{fig: prvsab}. It is the presence of the second term in Eq.(\ref{eq: calsdef}) which only depends on $\alpha_H$ to break the degeneracy allowing to constrain both $\alpha_H$ and $\beta_1$ separately instead of their combination in $\Xi_1$ only.


\section{Theoretical vs universal pressure profile}

Eq.(\ref{eq: prdhost}) is derived from the hydrostatic equilibrium equation under the assumptions of double\,-\,$\beta$ model for the electron density, and NFW for the dark halo mass density. Although reasonable, both are aprioristic hypotheses so that it is worth wondering how the resulting pressure profile compares with observed ones. This comparison can be carried out in the GR case since we are only interested in checking whether the theoretical $P(r)$ is reasonably in agreement with what is observed rather than fitting actual data. We can therefore set $(\alpha_H, \beta_1, \gamma_N) = (0, 0, 1)$ for the rest of this section.

\begin{table*}
\begin{center}
\resizebox{\textwidth}{0.41 \paperheight}{
\begin{tabular}{ccccccccccc} 
\hline 
id & $z$ & $\log{(M_{500}/M_{\odot})}$ & $c_{500}$ & $\log{(M_{200}/M_{\odot})}$ & $c_{200}$ & $\beta$ & $\log{x_{c1}}$ & $\log{x_{c2}}$ & $\log{(n_{01}/cm^{-3})}$ & $\log{(n_02/n_0)}$ \\
\hline \hline
Abell2204 & 0.15 & 15.01 & 2.52 & 15.18 & 3.87 & 0.641 & -2.07 & -1.23 & -0.64 & -1.17 \\
Abell383 & 0.19 & 14.67 & 2.71 & 14.83 & 4.14 & 0.601 & -1.98 & -1.33 & -0.87 & -0.84 \\
Abell1423 & 0.21 & 14.94 & 2.53 & 15.10 & 3.89 & 0.497 & -1.62 & ----- & -1.58 & ----- \\
Abell209 & 0.21 & 15.10 & 2.43 & 15.27 & 3.75 & 0.586 & -1.12 & ----- & -2.08 & ----- \\
Abell963 & 0.21 & 14.83 & 2.59 & 14.99 & 3.98 & 0.663 & -1.54 & -1.00 & -1.51 & -0.58 \\
Abell2261 & 0.22 & 15.16 & 2.39 & 15.33 & 3.69 & 0.581 & -1.87 & -1.35 & -1.34 & -0.45 \\
Abell2219 & 0.23 & 15.28 & 2.32 & 15.45 & 3.59 & 0.682 & -1.03 & ----- & -1.97 & ----- \\
Abell267 & 0.23 & 14.82 & 2.59 & 14.98 & 3.97 & 0.639 & -1.16 & ----- & -1.92 & ----- \\
RXJ21296 & 0.24 & 14.89 & 2.54 & 15.05 & 3.91 & 0.548 & -1.72 & ----- & -1.14 & ----- \\
Abell1835 & 0.25 & 15.09 & 2.42 & 15.26 & 3.73 & 0.669 & -1.86 & -1.15 & -0.85 & -0.99 \\
Abell697 & 0.28 & 15.23 & 2.32 & 15.40 & 3.59 & 0.639 & -1.09 & ----- & -1.99 & ----- \\
Abell611 & 0.29 & 14.87 & 2.52 & 15.03 & 3.88 & 0.597 & -2.28 & -1.30 & -0.84 & -0.87 \\
MS2137 & 0.31 & 14.67 & 2.62 & 14.83 & 4.03 & 0.491 & -2.03 & ----- & -0.94 & ----- \\
MACSJ1931 & 0.35 & 15.00 & 2.41 & 15.16 & 3.72 & 0.689 & -1.82 & -1.12 & -0.76 & -1.09 \\
AbellS1063 & 0.35 & 15.35 & 2.23 & 15.52 & 3.46 & 0.676 & -1.74 & -1.21 & -1.46 & -0.20 \\
MACSJ1115 & 0.36 & 14.93 & 2.44 & 15.10 & 3.76 & 0.647 & -1.70 & -1.11 & -1.04 & -0.86 \\
MACSJ1532 & 0.36 & 14.98 & 2.42 & 15.14 & 3.73 & 0.614 & -1.70 & ----- & -0.91 & ----- \\
Abell370 & 0.38 & 15.07 & 2.36 & 15.24 & 3.64 & 0.708 & -0.89 & ----- & -2.24 & ----- \\
ZWCL0024 & 0.39 & 14.64 & 2.59 & 14.80 & 3.97 & 0.453 & -1.47 & ----- & -1.83 & ----- \\
MACSJ1720 & 0.39 & 14.80 & 2.50 & 14.96 & 3.85 & 0.747 & -1.68 & -0.97 & -1.06 & -0.93 \\
MACSJ0429 & 0.40 & 14.76 & 2.51 & 14.93 & 3.87 & 0.669 & -1.80 & -1.07 & -0.87 & -1.05 \\
MACSJ2211 & 0.40 & 15.26 & 2.25 & 15.43 & 3.49 & 0.667 & -1.30 & ----- & -1.50 & ----- \\
MACSJ0416 & 0.42 & 14.96 & 2.39 & 15.13 & 3.69 & 1.104 & -0.64 & ----- & -2.24 & ----- \\
MACSJ0451 & 0.43 & 14.80 & 2.47 & 14.96 & 3.81 & 0.683 & -0.99 & ----- & -1.98 & ----- \\
MACSJ0417 & 0.44 & 15.34 & 2.19 & 15.52 & 3.40 & 0.709 & -1.76 & -0.85 & -1.05 & -1.13 \\
MACSJ1206 & 0.44 & 15.28 & 2.22 & 15.46 & 3.44 & 0.722 & -1.54 & -0.98 & -1.41 & -0.60 \\
MACSJ0329 & 0.45 & 14.90 & 2.41 & 15.06 & 3.72 & 0.749 & -1.74 & -0.93 & -0.89 & -1.16 \\
MACSJ1347 & 0.45 & 15.34 & 2.19 & 15.51 & 3.40 & 0.661 & -2.11 & -1.41 & -0.50 & -0.82 \\
MACSJ1311 & 0.49 & 14.59 & 2.55 & 14.75 & 3.92 & 0.925 & -1.24 & -0.76 & -1.40 & -0.75 \\
MACSJ0257 & 0.50 & 14.93 & 2.36 & 15.10 & 3.65 & 0.584 & -1.31 & ----- & -1.56 & ----- \\
MACSJ0911 & 0.50 & 14.95 & 2.35 & 15.12 & 3.63 & 0.557 & -1.07 & ----- & -2.08 & ----- \\
MACSJ2214 & 0.50 & 15.12 & 2.27 & 15.29 & 3.51 & 0.600 & -1.18 & ----- & -1.81 & ----- \\
MACSJ0018 & 0.54 & 15.22 & 2.20 & 15.39 & 3.42 & 0.703 & -1.59 & -0.95 & -1.79 & -0.20 \\
MACSJ1149 & 0.54 & 15.27 & 2.18 & 15.45 & 3.38 & 0.720 & -0.85 & ----- & -2.17 & ----- \\
MACSJ0717 & 0.55 & 15.40 & 2.11 & 15.57 & 3.29 & 1.003 & -1.30 & -0.66 & -1.87 & -0.33 \\
MACSJ1423 & 0.55 & 14.82 & 2.39 & 14.99 & 3.69 & 0.556 & -1.88 & ----- & -0.70 & ----- \\
MACSJ0454 & 0.55 & 15.06 & 2.27 & 15.23 & 3.52 & 0.631 & -1.09 & ----- & -1.75 & ----- \\
MACSJ0025 & 0.58 & 14.88 & 2.34 & 15.05 & 3.62 & 0.878 & -0.74 & ----- & -2.14 & ----- \\
MS2053 & 0.58 & 14.48 & 2.55 & 14.64 & 3.92 & 0.604 & -1.08 & ----- & -1.94 & ----- \\
MACSJ0647 & 0.59 & 15.04 & 2.26 & 15.21 & 3.50 & 0.636 & -1.17 & ----- & -1.69 & ----- \\
MACSJ2129 & 0.59 & 15.03 & 2.27 & 15.20 & 3.51 & 0.620 & -1.17 & ----- & -1.73 & ----- \\
MACSJ0744 & 0.69 & 15.10 & 2.18 & 15.27 & 3.39 & 0.622 & -1.76 & -1.13 & -1.06 & -0.72 \\
MS1054 & 0.83 & 14.95 & 2.17 & 15.13 & 3.38 & 1.168 & -0.48 & ----- & -2.17 & ----- \\
CLJ0152 & 0.83 & 14.89 & 2.20 & 15.07 & 3.41 & 1.717 & -0.17 & ----- & -2.59 & ----- \\
\hline
\end{tabular}}
\caption{Input BOXSZ \cite{Shitanishi:2017xct} cluster sample. We report the cluster id, the redshift, the halo mass and concentration for $\Delta = (500, 200)$, and the best fit parameters of the double\,-\,$\beta$ profile.}
\label{tab: inputsample}
\end{center}
\end{table*}

\begin{figure*}
\centering
\includegraphics[width=0.65\columnwidth]{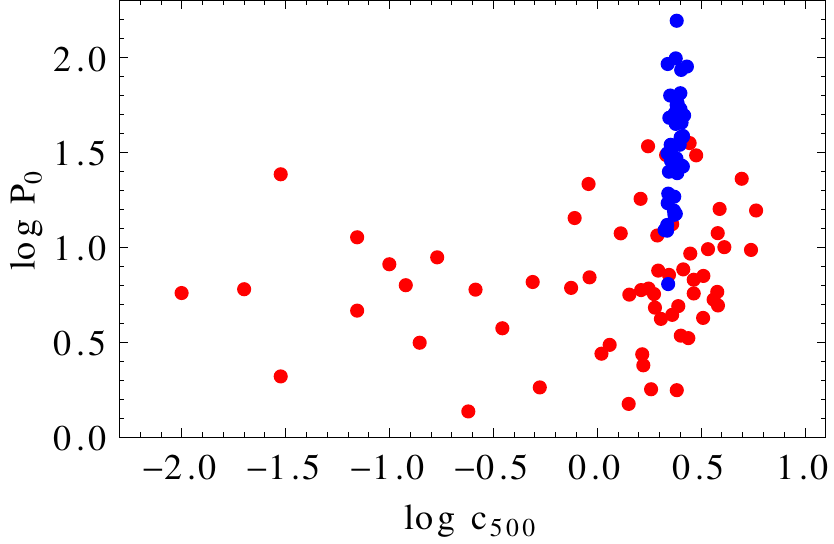}
\includegraphics[width=0.65\columnwidth]{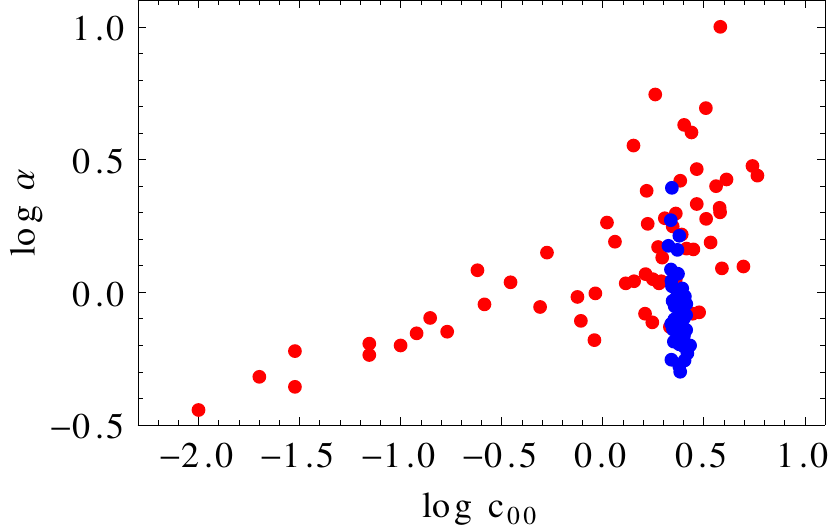}
\includegraphics[width=0.65\columnwidth]{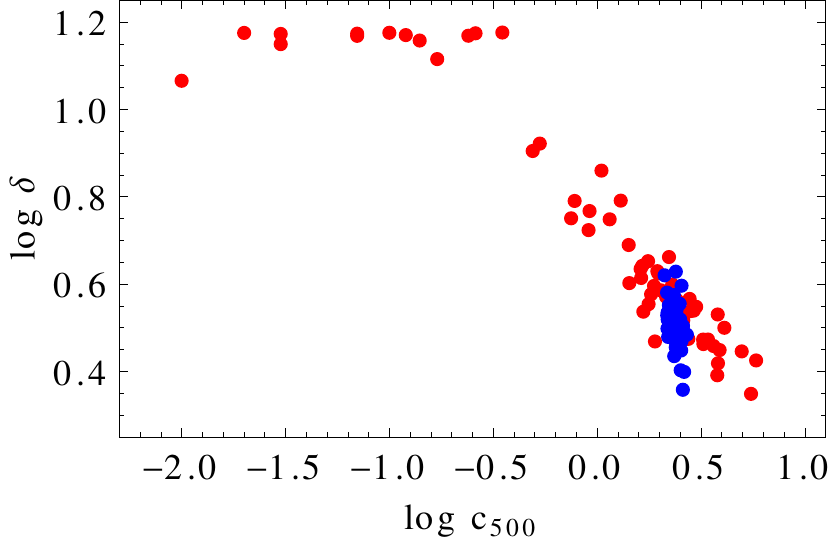} \\
\caption{Distribution of universal pressure profile parameters as inferred from the fit to the theoretical pressure profiles (blue) and to Planck clusters (red). Note that we show logarithm to improve the dynamic range of the plots.}
\label{fig: fitup}
\end{figure*}

To this end, we compute $P(r)$ over the range $(0.1, 1.5) R_{500}$, and fit it with the universal pressure profile \cite{Nagai:2007mt,Arnaud:2009tt} given by

\begin{equation}
P(r) = \frac{P_0 P_{500}}
{(c_{500} r/R_{500})^{\gamma}
[1 + (c_{500} r/R_{500})^{\alpha}]^{(\delta - \gamma)/\alpha}}
\label{eq: pup}
\end{equation}
with $(P_0, \alpha, \delta, \gamma)$ fitting parameters\footnote{Note that, in the literature, $\delta$ in Eq.(\ref{eq: pup}) is denoted as $\beta$ but we have changed here the notation to avoid confusion with the $\beta$ parameter of the electron density profile.}, and $P_{500}$ a redshift and mass dependent normalization which, following \cite{Ade:2012ngj}, we set as\footnote{Note that the exponent of the mass term deviates from the self\,-\,similar scaling $2/3$ by an additional 0.12 this allows a better match of the stacked profiles.}

\begin{equation}
P_{500} = 1.65 \times 10^{-3} \left ( \frac{M_{500}}{3 \times 10^{14} h_{70}^{-1} M_{\odot}} \right )^{\frac{2}{3} + 0.12} E^{8/3}(z) h_{70}^{2}  \ ,
\label{eq: p500def}
\end{equation}
with $P_{500}$ in ${\rm keV/cm^3}$, $h_{70}$ the Hubble constant $H_0$ in units of $70 \ {\rm km/s/Mpc}$, and $E^2(z) = \Omega_M (1 + z)^3 + (1 - \Omega_M)$ for the $\Lambda$CDM model we take as background for the normalization of the parametric pressure profile (setting $\Omega_M = 0.32$ and $H_0 = 67 \ {\rm km/s/Mpc}$ in accordance with Planck cosmological results \cite{Aghanim:2018eyx}).

The ${\cal{Q}}_{0}^{GR}(r)$ function entering the theoretical pressure profile in Eq.(\ref{eq: prdhost}) depends on the values of the double\,-\,$\beta$ model parameters. For later applications, it is convenient to reparameterize them as 

\begin{equation}
\begin{array}{l}
\displaystyle{\{\beta, r_{c1}, r_{c2}, n_{01}, n_{02} \}} \ \longrightarrow \\  
\displaystyle{\{ \beta, \log{x_{c1}}, \log{x_{c2}}, \log{n_{01}}, \log{(n_{02}/n_{01})} \}} \\
\end{array}
\label{eq: dbpar}
\end{equation}
with $x_{ci} = r_{ci}/R_{200}$. To set these quantities, we consider the cluster sample assembled by the BOXSZ project\footnote{We remove one object from the BOXSZ sample since it has an anomalously large $\beta$ (14.28 vs a typical value $\beta \sim 0.6)$ so that we are left with 44 clusters.} \cite{Shitanishi:2017xct} since this spans a quite wide range in both mass $(3.0 \le M_{500}/10^{14} {\rm M_{\odot}} \le 25)$ and redshift $(0.15 \le z \le 0.83)$. We convert the values of $(\beta, r_{c1}, r_{c2}, n_{01}, n_{02})$ in their paper to the parameters in Eq.(\ref{eq: dbpar}) after converting the reported $M_{500}$ into $M_{200}$ under the assumption of NFW model and setting $c_{200}$ according to the mass\,-\,concentration relation in \cite{Dutton:2014xda}. The parameters thus obtained are given in Table\,\ref{tab: inputsample} where a sign ----- is given for $\log{x_{c2}}$ and $\log{(n_{02}/n_{01})}$ if the cluster is better fitted by a single\,-\,$\beta$ profile. Strictly speaking, the use of a mass\,-\,concentration relation to set $c_{200}$ before converting from $M_{500}$ to $M_{200}$ and set $c_{200}$ implicitly assumes that Newtonian gravity holds since the $M_{200}$\,-\,$c_{200}$ relation has been inferred from N\,-\,body simulations under this framework. However, we here just want to have realistic profiles for both the convergence and the pressure which is what we indeed get in this way. 

We use the parameters in Table\,\ref{tab: inputsample} to generate the theoretical pressure profile and fit it with Eq.(\ref{eq: pup}) adjusting the parameters $(P_0, \alpha, \delta)$ while keeping $c_{500}$ to the input value and fixing $\gamma = 0.31$ as recommended in \cite{Ade:2012ngj}. The quality of the fit can be guessed computing 

\begin{equation}
\varepsilon_{rms} = 2 \sqrt{\left \langle 
\left [ \frac{P_{UP}(R) - P_{th}(R)}{P_{UP}(R) + P_{th}(R)}
\right ]^2 \right \rangle}
\label{eq: defepsrms}    
\end{equation}
where $P_{UP}(R)$ and $P_{th}(R)$ are the universal profile and the theoretical one, and the mean is taken over the range $(0.1, 1.5) R_{500}$ sampled in steps of $0.01 R_{500}$. For the 44 clusters in the sample, we find quite small $\varepsilon_{rms}$ values with $\varepsilon_{rms} = 2.06\%$ as median, and a $95\%$\,CL spanning the range $(0.05, 5.23)\%$. Although these numbers tell us that the universal pressure profile excellently fits the theoretical one, they are not enough to judge whether the shapes are realistic. We must rather compare the value of the fitting parameters to those obtained fitting real clusters. To this aim, we can rely on the values reported in \cite{Ade:2012ngj} where the fit has been performed for the subsample of 62 clusters with SZ measurements from the early Planck data release. The median, 68 and $95\%$\,CL of the best fit parameters to the theoretical BOXSZ and observed Planck clusters are as follows

\begin{displaymath}
\begin{array}{l}
P_0\,: \ \ 34_{-18 \ -22}^{+29 \ +65} \ \ {\rm vs} \ \ 6.1_{-3.0 \ -4.6}^{+9.5 \ +27.8} \ \ ; \\
\alpha\,: \ \ 0.81_{-0.18 \ -0.29}^{+0.36 \ +1.06} \ \ {\rm vs} \ \ 1.24_{-0.53 \ -0.80}^{+1.42 \ +4.32} \ \ ; \\
\delta\,: \ \ 3.23_{-0.39 \ -0.73}^{+0.33 \ +0.93} \ \ {\rm vs} \ \ 3.99_{-1.02 \ -1.53}^{+10.4 \ +11.0} \ \ . \\
\end{array}
\end{displaymath}
Although the ranges have a good overlap, we can nevertheless note that the typical $P_0$ values for our sample are definitely larger than those for Planck clusters. Moreover, our $\alpha$ values are smaller, and we do not find systems with extremely large $\delta$ values. These differences likely originate by our choice of taking $c_{500}$ fixed to the value inferred from the mass\,-\,concentration relation. On the contrary, the fit performed in \cite{Ade:2012ngj} considers $c_{500}$ as a parameter to be optimised. In particular, also values smaller than $1$ are allowed so that one can get halos with $c_{200} < 1$ which, although mathematically possible, it has no physical sense given that one would get $R_{200} < r_s$. As a consequence, the two samples of clusters have a radically different distribution of $c_{500}$ values, the $95\%$\,CL ranges being $(2.17, 2.62)$ vs $(0.02, 5.51)$. The correlation between $c_{500}$ and the other fit parameters then motivates the discrepancy between the results for the Planck clusters and our theoretical pressure profiles. This can be seen in Fig.\,\ref{fig: fitup} where we show the distribution of the fit parameters for the two samples. As it is evident, the Planck clusters cover a much larger range in $c_{500}$ including unrealistically small values. On the contrary, $c_{500}$ is fixed for the theoretical pressure hence spanning a definitely smaller range because of the use of the mass\,-\,concentration relation. However, over the $c_{500}$ range in common between the two samples the $(P_0, \alpha, \delta)$ values are comparable, our sample lacking systems with unusually large $(\alpha, \delta)$ values which are needed to compensate for the unrealistically small $c_{500}$. This result shows that our assumptions on the electron density and dark halo mass profiles and the use of the hydrostatic equilibrium equation generate theoretical pressure profiles which share the same properties of a subsample of real clusters detected by Planck. This is a reassuring evidence that we can rely on the present modeling to describe realistic systems.                                         

\section{Fisher matrix forecast}

The different ways the parameters $(\alpha_H, \beta_1, \gamma_N)$ enter the convergence and pressure profiles and the possibility to constrain the electron density parameters through X\,-\,ray measurements make it intriguing to wonder whether a joint use of the three probes\footnote{Although the electron density profile does not depend on the DHOST parameter, it helps to constrain the double\,\-,$\beta$ model ones hence indirectly improving the determination of $(\alpha_H, \beta_1, \gamma_N)$ thanks to the effective priors it imposes on the astrophysical ones.} can significantly constrain DSHOT theories. A quick way to investigate this issue is to rely on Fisher matrix forecasts. Under the assumption that the likelihood ${\cal{L}}({\bf p})$ is Gaussian for a given observable ${\cal{O}}$, one can estimate the covariance matrix of the model parameters ${\bf p}$ inverting the Fisher matrix whose elements are given by

\begin{equation}
F_{\alpha \beta} = - \left <  \frac{\partial^2 \ln{{\cal{L}}({\bf p})}}
{\partial p_{\alpha} p_{\beta}} \right >
= \frac{\partial {\bf D }({\bf p})}{\partial p_{\alpha}} 
{\bf C}_{obs}^{-1} \frac{\partial {\bf D}({\bf p})}{\partial p_{\beta}}
\label{eq: fabdef}
\end{equation}
where in the second equality we have used the Gaussian approximation of the likelihood, and denoted with ${\bf D}({\bf p})$ the theoretical data vector (i.e., the vector whose elements are the predicted values of the observable ${\cal{O}}$ estimated at the position where data are available) and ${\bf C}_{obs}$ is the data covariance matrix. Note that, because of the ergodic principle, we have replaced the spatial average with the evaluation for the fiducial model parameters. If more than one observable is available and they are statistically independent, the total Fisher matrix is the sum of those for each probe, while a prior on the model parameters (from theoretical principle or other measurements) can be added as a diagonal matrix with the inverse of variance as diagonal elements. Finally, according to the Cramer\,-\,Rao inequality, the best constraints one can obtain on the model parameters ${\bf p}$ are given by the diagonal elements of the inverse of the Fisher matrix, while off diagonal elements can be used to quantify the correlation among parameters.

\subsection{Convergence Fisher matrix}

As a first probe, we consider the convergence profile $\kappa(R)$ assuming that it is measured in ${\cal{N}}_{WL}$ logarithmically equispaced radial points over the angular range $(\theta_{min}, \theta_{max})$. Following common approach, we assume the errors on the measurement are uncorrelated so that the error covariance matrix is simply diagonal. The Fisher matrix elements are then given by 

\begin{equation}
F_{\alpha \beta}^{WL} =\sum_{i = 1}^{{\cal{N}}_{WL}}
{\frac{1}{\sigma_{\kappa}^2(\theta_i)}
\frac{\partial \kappa(\theta_i)}{\partial p_{\alpha}} 
\frac{\partial \kappa(\theta_i)}{\partial p_{\beta}}} \ ,
\label{eq: kappafm}
\end{equation}
where we remind the reader the relation $\theta = R \times [(206265/60)/ D_L(z)]$ converts the linear distance $R$ (in kpc) from the centre to the angular ones $\theta$ (in arcmin) given a cosmological model for the estimate of the angular diameter distance $D_L(z)$. One can rearrange Eq.(\ref{eq: kappafm}) as follows\,:

\begin{eqnarray}
F_{\alpha \beta}^{WL} & = & \sum_{i = 1}^{{\cal{N}}_{WL}}{
\left [ \frac{\kappa(\theta_i)}{\sigma_{\kappa}(\theta_i)} \right ]^2
\frac{1}{\kappa(\theta_i)} \frac{\partial \kappa(\theta_i)}{\partial p_{\alpha}}
\frac{1}{\kappa(\theta_i)} \frac{\partial \kappa(\theta_i)}{\partial p_{\beta}}} \nonumber \\ 
 & = & \sum_{i = 1}^{{\cal{N}}_{WL}}{
\nu_{\kappa}^{2}(\theta_i) 
\frac{\partial \ln{\kappa(\theta_i)}}{\partial p_{\alpha}}
\frac{\partial \ln{\kappa(\theta_i)}}{\partial p_{\beta}}}
\label{eq: kappafmbis}
\end{eqnarray}
where $\nu_{\kappa}(\theta_i)$ is the S/N ratio of the convergence measured in the position $\theta_i$. In order to approximately model this quantity, we follow a method similar to the one in \cite{Cardone:2016jdt} which we refer the reader to for details. Here, we only sketch the procedure. We first compute the S/N including both the measurement error $\epsilon_{\kappa}(\theta)$ and the systematic floor due to the ellipticity intrinsic dispersion (which we set as $\sigma_e = 0.22$). This gives

\begin{equation}
\nu_{\kappa}(\theta) = \frac{{\cal{W}}_\kappa \kappa_{\infty}(\theta)}{\epsilon_{\kappa}(\theta)}
\left [ 1 + \left ( \frac{\sigma_e}{\epsilon_{\kappa}} \right )^2
\frac{1}{2 n_g {\cal{A}}} \right ]^{-1/2}
\label{eq: stonkappa}
\end{equation}
where $\kappa_{\infty}$ is the convergence evaluated for a source at infinity, and ${\cal{W}}_\kappa$ is a dilution factor accounting for the redshift distribution of the sources. The term $2 n_g {\cal{A}}$ downgrades the systematic floor depending on the number of sources which is computed using the number density $n_g$ and area ${\cal{A}}$ of the circular corona centred on $\theta$. We consider that measurements will be performed by a Euclid\,-\,like experiment \cite{Laureijs:2011gra} hence we adopt the Euclid redshift distribution in \cite{Blanchard:2019oqi} and set $n_g = 30 \ {\rm gal/arcmin}^2$. We use Eq.(\ref{eq: stonkappa}) to compute the S/N for a large number of $(\theta, \log{M_{200}}, z, \epsilon/\kappa)$ values and then fit the median results to get an approximate scaling of the S/N with the angular distance, the halo mass, and the redshift. 

We use this approximated relation as input to Eq.(\ref{eq: kappafmbis}) where we also fix $(\theta_{min}, \theta_{max}) = (0.2, 10.2) \ {\rm arcmin}$ for all clusters, no matter their redshift. Note that actual observations may probe a still larger angular range with $\theta_{max} \sim 16 \ {\rm arcmin}$ for the CLASH sample \cite{Umetsu:2014vna} which covers a similar redshift range as BOXSZ. We have preferred to be conservative cutting the upper limit should the clusters one finally adopts be smaller in angular size than the CLASH ones.

For a given cluster, the model parameters can be split in two groups. On one hand, we have the astrophysical ones which are specific of that cluster. These are the halo mass\footnote{Following common practice, we use the logarithm of the halo mass rather than the mass itself in order to explore a wider range dealing with order unity quantities. Moreover, with an abuse of notation, we denote with $M_{200}$ the mass in solar units $M_{\odot}$ so we drop the $M_{\odot}$ from $\log{(M_{200}/M_{\odot})}$.} $\log{M_{200}}$ and the concentration $c_{200}$ of the NFW halo. We avoid to add a prior on the concentration based on the mass\,-\,concentration relation derived from N\,-\,body simulations because it has been obtained postulating Newtonian gravity\footnote{Such a choice could look contradictory given that we have used such a relation to set the concentration for the fiducial cluster parameters. However, in that case, we were only interested in obtaining realistic profiles and the use of the mass\,-\,concentration relation ensured that this goal is achieved. When dealing with actual data, one will not make any assumption on the $c_{200}$\,-\,$M_{200}$ relation so that we do not include it in our forecast.}. The remaining parameters are the DHOST ones $(\alpha_H, \beta_1, \gamma_N)$ which are universal quantities, but redshift dependent. It is therefore important to stress that, although we do not explicitly denote it to shorten the notation, what the data are constraining are $(\alpha_H, \beta_1, \gamma_N)$ at the lensing cluster redshift $z$. As a consequence, one can not stack together all the clusters in a given sample so that we compute Fisher matrix forecasts from individual convergence profiles. For completeness, we report in Appendix B.1 the derivatives needed for the Fisher matrix computation. 

\subsection{Pressure profile Fisher matrix}

We assume to use the SZ data to sample the pressure profile $P(r)$ in ${\cal{N}}_{SZ}$ linearly spaced radial distance $r_i$ from cluster centre over the range $(\xi_{min}, \xi_{max}) R_{500}$. Neglecting any correlation among the errors, the pressure profile Fisher matrix may then be written as

\begin{eqnarray}
F_{\alpha \beta}^{SZ} & = & \sum_{i = 1}^{{\cal{N}}_{SZ}}
{\frac{1}{\sigma_{P}^2(r_i)}
\frac{\partial P(r_i)}{\partial p_{\alpha}} 
\frac{\partial P(r_i)}{\partial p_{\beta}}} \nonumber \\ 
& = &
\sum_{i = 1}^{{\cal{N}}_{SZ}}
{\nu_{P}^2(r_i) \frac{\partial \ln{P(r_i)}}{\partial p_{\alpha}}  
\frac{\partial \ln{P(r_i)}}{\partial p_{\beta}}} \ ,
\label{eq: pressfm}
\end{eqnarray}
where we have made the same rearrangement of the terms as for WL, but for the SZ data denoting with $\nu_P(r)$ the S/N of pressure measurements.

Appendix B.2 reports the relevant derivatives entering the pressure Fisher matrix, while the S/N is approximated as a function of the dimensionless distance $\xi = r/R_{500}$ and the mass $\log{M_{500}}$ using the pressure profile inferred by SZ data of the X\,-\,COP sample \cite{Ghirardini:2018byi}. It is worth noting that the X\,-\,COP clusters span a comparable mass range, but at a much smaller redshift ($z \le 0.09$). Although the SZ signal is independent on $z$, the precision on the measurements can depend on the size of the cluster for a given angular resolution of the instrument. As such, the S/N for BOXSZ\,-\,like clusters could be different from what we have inferred from the X\,-\,COP sample. To account for this difference, we will introduce later a correction factor which allows us to investigate the impact of deviations of the the actual $\nu_P(r)$ from the one assumed here.

The probed radial range is fixed based on the following considerations. First, we remove the very inner part which can be affected by deviations from hydrostatic equilibrium so that we set $\xi_{min} = 0.1$ as conservative limit. We then look at the radial extent of the BOXSZ data finding $\xi_{max} = 1.17$ as median value, $0.87 \le \xi_{max} \le 1.81$ as $68\%$\,CL. We therefore compute the pressure Fisher matrix for three different cases denoted as {\it central}, {\it intermediate}, {\it large range} with $\xi_{max} = (0.87, 1.17, 1.81)$, respectively.

It is worth noting that the list of model parameters at play now is definitely larger than for the convergence. While the DHOST parameters $(\alpha_H, \beta_1, \gamma_N)$ are still the same, there is a larger number of astrophysical parameters to be marginalized over. Indeed, beside the halo mass $\log{M_{200}}$ and the concentration $c_{200}$ of the NFW halo, we now have to set the parameters of the double\,-\,$\beta$ model given in Eq.(\ref{eq: dbpar}). The total number of parameters is, therefore, 10 reducing to 8 for objects better fit by the single\,-\,$\beta$ model. As a consequence, we do not expect pressure data alone to be able to put meaningful constraints on $(\alpha_H, \beta_1, \gamma_N)$, but they are nevertheless of utmost importance thanks to the possibility of breaking degeneracy.

\subsection{Electron density Fisher matrix}

The gas density plays a different role than convergence and pressure in the present analysis. Indeed, we do not compute it from a theoretical model, but directly fit an empirical profile to the electron density data as measured from X\,-\,ray (hereafter XR) data. The corresponding Fisher matrix may be simply computed as 
\begin{equation}
F_{\alpha \beta}^{XR} =  
\sum_{i = 1}^{{\cal{N}}_{XR}}
{\nu_{e}^2(r_i) \frac{\partial \ln{n_e(r_i)}}{\partial p_{\alpha}}  
\frac{\partial \ln{n_e(r_i)}}{\partial p_{\beta}}} \ ,
\label{eq: gasfm}
\end{equation}
where we have directly used the formulation with the S/N $\nu_e(r)$ highlighted, and the sum is over ${\cal{N}}_{XR}$ measured points linearly spaced over the range $(\xi_{min}, \xi_{max}) R_{500}$. We use the same $(\xi_{min}, \xi_{max})$ values adopted for the pressure Fisher matrix, and still rely on X\,-\,COP data to infer the electron density S/N as function of $\xi$ and $\log{M_{500}}$. 

It is worth noticing that the electron density does not depend on the DHOST theory parameters so that X\,-\,ray data are unable to directly constrain these quantities. They are nevertheless of great help since they strongly constrain the parameters which input the pressure profile. 

\subsection{Fiducial model parameters and Fisher matrix setup}

In order to compute the Fisher matrix, for each given cluster, there are some parameters and choices that have to be made. First, we need to set the cluster redshift $z$, halo mass and concentration $(\log{M_{200}}, c_{200})$, and the double\,-\,beta model parameters $\{\beta, \log{x_{c1}}, \log{x_{c2}}, \log{n_{01}}, \log{(n_{02}/n_{01})}\})$. These are taken from the values in Table\,\ref{tab: inputsample} which are then used as input to the convergence, pressure, and electron density Fisher matrices which we will hereafter refer to as WL, SZ, and XR, respectively. We have, however, to set also the DHOST theory parameters $(\alpha_H, \beta_1, \gamma_N)$. We will only consider models with $G_{eff}^{N} = G_N$ hence setting $\gamma_N = 1 - \alpha_H - 3 \beta_1$. We stress that this assumption is only made to reduce the arbitrariness in the choice of the fiducial DHOST parameters, but in the analysis we do not impose it so that $\gamma_N$ is still a quantity to constrain. 

\begin{table}
\begin{center}
\begin{tabular}{cccccc} 
\hline 
id & $\alpha_H$ & $\beta_1$ & $\Xi_1$ & $\Xi_2$ & $\Xi_3$ \\
\hline \hline
GR & 0.0 & 0.0 & 0.0 & 0.0 & 0.0 \\
\hline
A0M & 0.0 & -0.15 & 0.0375 & 0.0 & 0.0375 \\
A0P & 0.0 & 0.15 & -0.0375 & 0.0 & -0.0375  \\
\hline
B0M & -0.25 & 0.0 & 0.125 & -0.25 & 0.0 \\
B0P & 0.25 & 0.0 & -0.125 & 0.25 & 0.0 \\
\hline
D1M & 0.15 & -0.15 & 0.0 & 0.15 & 0.0 \\
D1P & -0.15 & 0.15 & 0.0 & -0.15 & 0.0 \\
\hline
D2M & 0.45 & -0.15 & -0.30 & 0.45 & 0.30 \\
D2P & -0.45 & 0.15 & 0.30 & -0.45 & -0.30 \\
\hline
\end{tabular}
\caption{DHOST fiducial models id, $(\alpha_H, \beta_1)$ parameters, and amplitudes of the terms setting deviations from GR. For all cases, it is $\gamma_N = 1 - \alpha_H - 3\beta_1$.}
\label{tab: dhostfid}
\end{center}
\end{table}

We are left with the two parameters $(\alpha_H, \beta_1)$ to set. The first obvious choice is the GR one, i.e., $(\alpha_H, \beta_1) = (0, 0)$. In this case, our analysis will tell us to which extent the data are able to discriminate between GR and DHOST based on how small are the constraints on $(\alpha_H, \beta_1, \gamma_N)$. It is also interesting, however, to investigate how the constraints change depending on the input fiducial. Indeed, some choices of the parameters may strengthen or weaken the deviations from GR hence making it easier or harder to spot them and constraining the parameters themselves. For this reason, we select other eight representative cases whose labels and parameters are summarised in Table\,\ref{tab: dhostfid}. Let us motivate their choice below. First class is obtained by setting $\alpha_H = 0$ so that it is 

\begin{displaymath}
\alpha_H = 0 \ \longrightarrow \ (\Xi_1, \Xi_2, \Xi_3) = (-\beta_1/4, 0, -\beta_1/4) \ . 
\end{displaymath}
Being $\Xi_2 = 0$, only one of the two corrective terms to the lensing convergence are present hence weakening the WL constraining power. A similar effect is obtained for the second class defined by the condition

\begin{displaymath}
\beta_1 = 0 \ \longrightarrow \ (\Xi_1, \Xi_2, \Xi_3) = (-\alpha_H/2, -\alpha_H, 0) 
\end{displaymath}
so that it is now the second corrective term to convergence to disappear. 

The pressure profile stays the same as GR for DHOST models with 

\begin{displaymath}
\alpha_H = -\beta_1 \ \longrightarrow \ (\Xi_1, \Xi_2, \Xi_3) = (0, -\beta_1, 0)
\end{displaymath}
which minimizes the impact of deviations from GR also for the convergence. On the contrary, both effects are maximized if we set 

\begin{displaymath}
\alpha_H = -3 \beta_1 \ \longrightarrow \ (2 \beta_1, -3 \beta_1, -2 \beta_1)
\end{displaymath}
which is the last class we consider. 

For all cases, we have to choose a value for $\alpha_H$ or $\beta_1$. Unfortunately, present day constraints on $(\alpha_H, \beta_1, \gamma_N)$ are still quite poor so that we must rather rely on theoretical motivations. In particular, Karmakar et al. (2019) have recently carried out an extensive analysis to set theoretical limits on the possible values of $(\alpha_H, \beta_1)$ at different redshifts. It turns out that, depending on $z$, $(\alpha_H, \beta_1)$ can only span a certain range. We choose the extreme values over the redshift range spanned by our data hence setting $\beta_1 = \pm 0.15$ for the first, second, and fourth class of models, and $\alpha_H = \pm 0.25$ for the third one. The models thus obtained are labeled as in Table\,\ref{tab: dhostfid}.

We have now all the parameters which are necessary as input to the computation of the WL, SZ, and XR Fisher matrices. However, we still need to set some quantities related to the observations. In particular, we must fix the number of WL, SZ, and XR measured points, i.e. $({\cal{N}}_{WL}, {\cal{N}}_{SZ}, {\cal{N}}_{XR}).$ We set ${\cal{N}}_{XR} = 40$ as a typical value inferred from X-COP data, while we investigate cases with 10 and 20 points in the WL and SZ datasets, having set ${\cal{N}}_{WL} = {\cal{N}}_{SZ}$ just to reduce the number of possible configurations to explore. A cautionary remark is in order for the S/N of the different data. Although our scaling relations are well motivated and based on actual data, it is nevertheless worth wondering how the constraints change with the S/N itself. Since we have based our WL S/N on a Euclid\,-\,like experiment whose features are well known, we do not change the WL S/N, while we allow for deviations of $\nu_P(r)$ and $\nu_e(r)$ from our assumptions. To this end, we modify Eqs.(\ref{eq: pressfm}) and (\ref{eq: gasfm}) by the following qualitative replacement

\begin{displaymath}
\nu_{P}(r) \ \longrightarrow \ {\cal{B}}_P \nu_{P}(r) \ \ , \ \ 
\nu_{e}(r) \ \longrightarrow \ {\cal{B}}_e \nu_{e}(r) \ \ ,  
\end{displaymath}
where $({\cal{B}}_P, {\cal{B}}_e)$ change the amplitude of the S/N, but not their radial profile (so that both S/N profile are still decreasing function of the distance). This is a simplifying assumption which guarantees enough flexibility for the aims of the present work.


\begin{figure*}
\centering
\includegraphics[width=0.65\columnwidth]{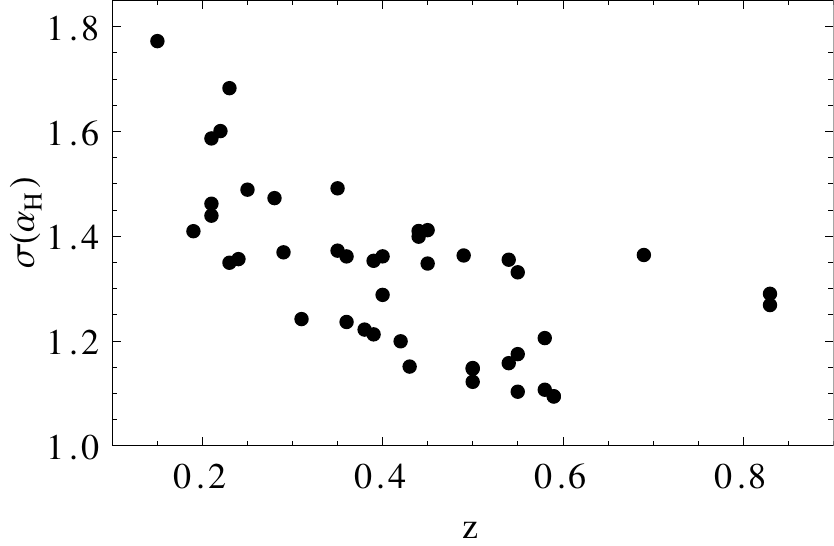}
\includegraphics[width=0.65\columnwidth]{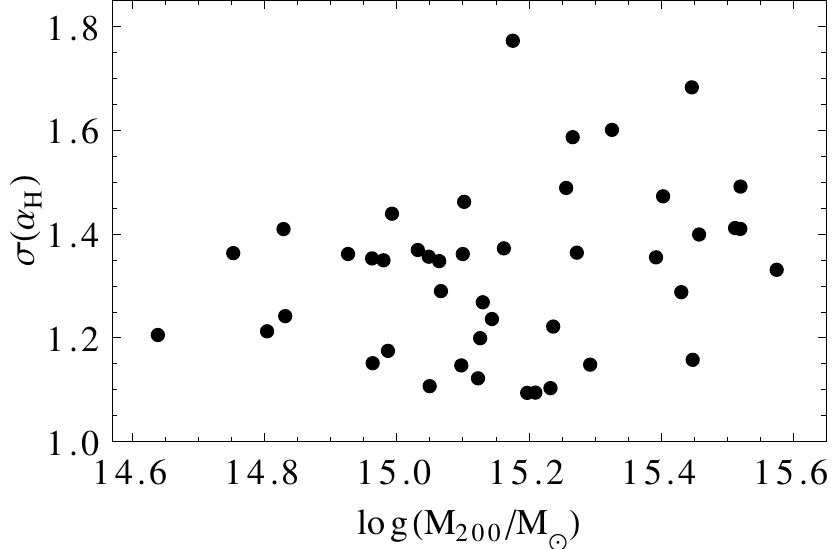}
\includegraphics[width=0.65\columnwidth]{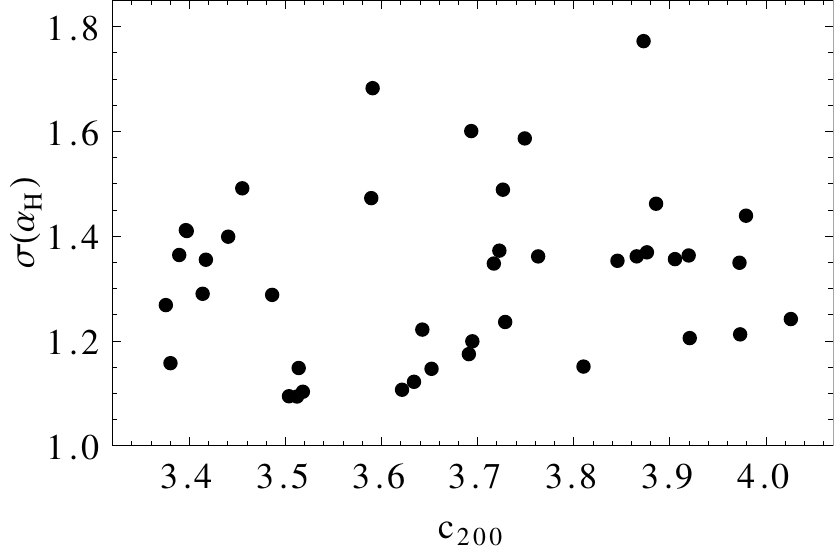} \\
\includegraphics[width=0.65\columnwidth]{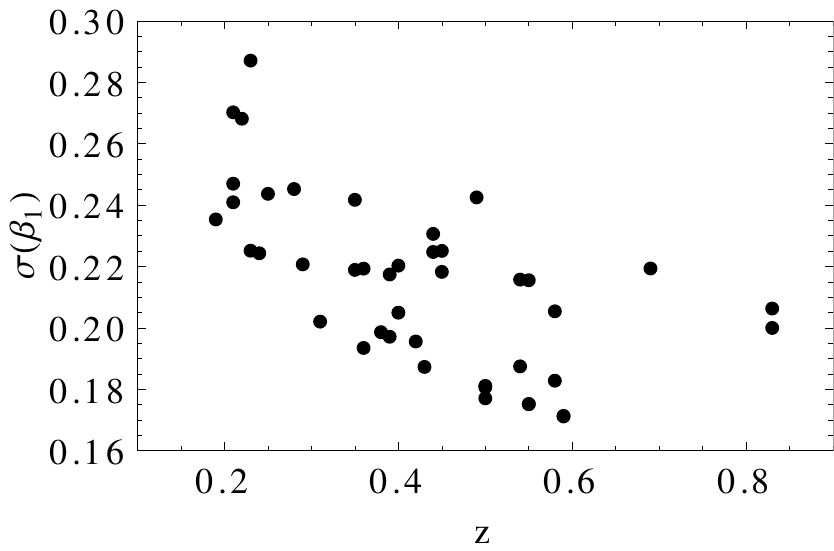}
\includegraphics[width=0.65\columnwidth]{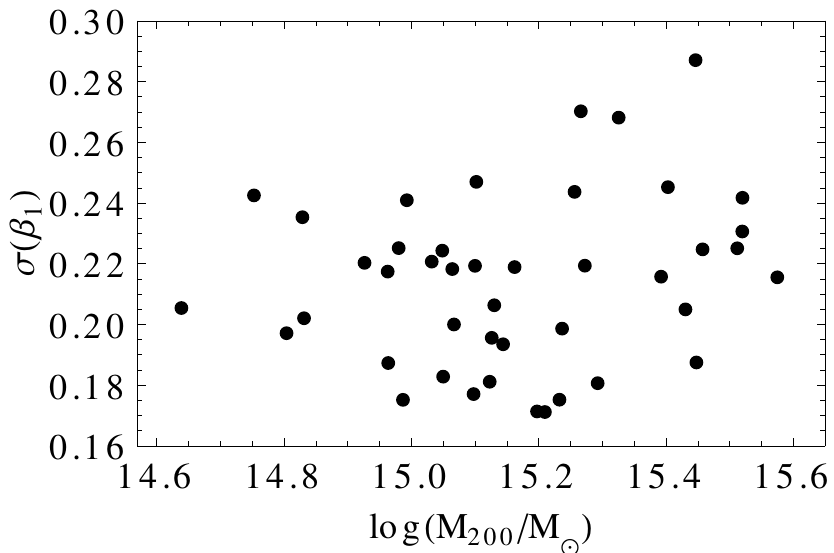}
\includegraphics[width=0.65\columnwidth]{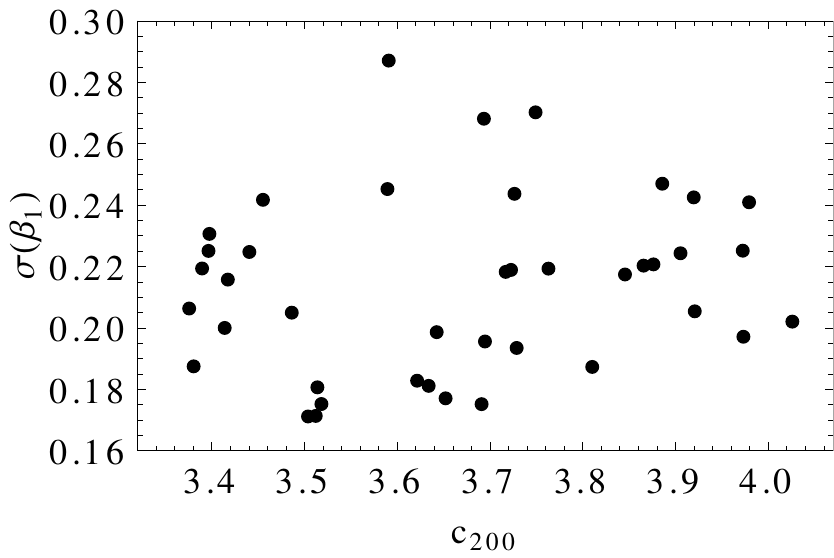} \\
\includegraphics[width=0.65\columnwidth]{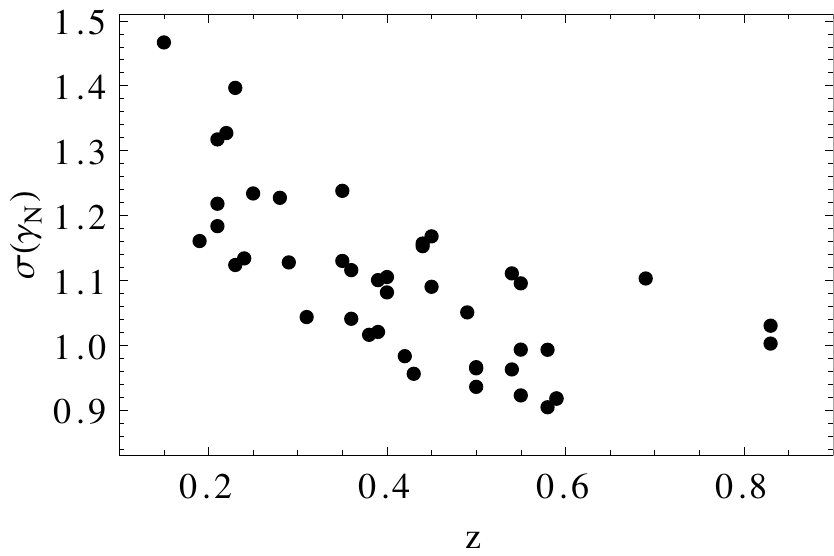}
\includegraphics[width=0.65\columnwidth]{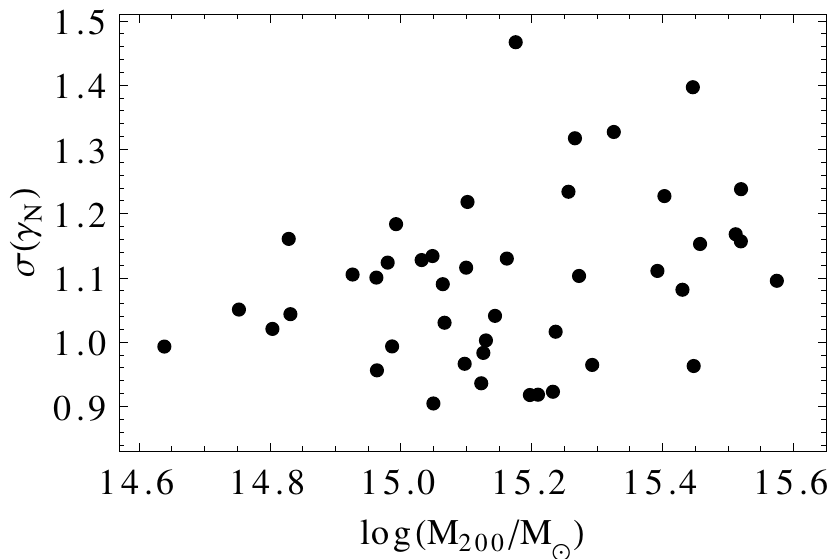}
\includegraphics[width=0.65\columnwidth]{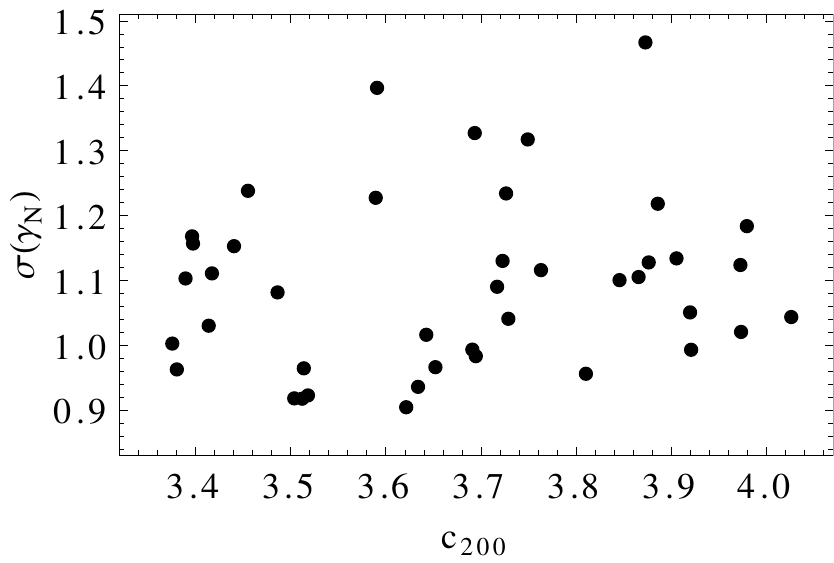} \\
\caption{Forecast errors on $(\alpha_H, \beta_1,\gamma_N)$ as a function of cluster redshift, mass, and concentration for the reference case, and the GR fiducial model.}
\label{fig: fmgrref} 
\end{figure*}

\section{Results}

We discuss below the constraints on the DHOST parameters we get from the Fisher matrix forecasts with the setup described above. We always marginalize over the NFW $\{\log{M_{200}}, c_{200}\}$, and the double\,-\,$\beta$ model $\{\beta, \log{x_{c1}}, \log{x_{c2}}, \log{n_{01}}, \log{(n_{02}/n_{01})}\}$ parameters since we are interested in constraining deviations from GR. We also remind the reader that, being the three datasets independent, the total Fisher matrix is just the sum of the three individual ones, i.e., $F_{tot} = F^{WL} + F^{SZ} + F^{XR}$. When investigating the dependence of the results on the assumption about the data, we will look at the ratio of the constraints on each single parameter with respect to those from an arbitrary chosen reference case. This is obtained setting $({\cal{N}}_{WL}, {\cal{N}}_{SZ}, {\cal{N}}_{XR}) = (10, 10, 40)$, $({\cal{B}}_P, {\cal{B}}_e) = (1.0, 1.0)$, and the {\it intermediate} case for the radial range probed by SZ and X\,-\,ray data. 

\subsection{Fiducial GR}

Let us first discuss the constraints on $(\alpha_H, \beta_1, \gamma_N)$ when GR is taken as fiducial. Not surprisingly, one gets almost no constraint at all if only pressure data are used because of the degeneracy among $(\alpha_H, \beta_1)$ which only enters combined into $\Xi_1$ and the presence of a up to 10 total parameters. Lensing convergence, on the contrary, works better thanks to the dependence on both $\Xi_2$ and $\Xi_3$ which allows to partially break the $(\alpha_H, \beta_1)$ degeneracy. Moreover, the lower number of nuisance parameters (only two) reduces overall the effect of marginalization. It is the joint use of WL, SZ, and XR datasets which significantly strenghten the constraints. Denoting with $\sigma(p_{\mu})$ the forecast error on the parameter $p_{\mu}$ and considering its distribution over the cluster sample, we get for the median, the 68 and 95$\%$\,CL

\begin{displaymath}
\begin{array}{l}
\sigma(\alpha_H) = 2.51 \ \ , \ \ (2.30, 2.71) \ \ , \ \ (2.11, 3.00) \ \ , \\
\sigma(\beta_1) = 0.77 \ \ , \ \ (0.43, 1.35) \ \ , \ \ (0.36, 2.13) \ \ , \\
\sigma(\gamma_N) = 5.85 \ \ , \ \ (4.70, 8.25) \ \ , \ \ (4.23, 10.5) \ \ , \\
\end{array}
\end{displaymath}
using lensing convergence only, which reduces to 

\begin{displaymath}
\begin{array}{l}
\sigma(\alpha_H) = 1.35 \ \  , \ \ (1.15, 1.47) \ \ , \ \ (1.09, 1.68) \ \ , \\
\sigma(\beta_1) = 0.22 \ \ , \ \ (0.18, 0.24) \ \ , \ \ (0.17, 0.29) \ \ , \\
\sigma(\gamma_N) = 1.10 \ \ , \ \ (0.96, 1.23) \ \ , \ \ (0.92, 1.40) \ \ , \\
\end{array}
\end{displaymath}
when WL\,+\,SZ\,+\,XR data are used. The improvement is particularly evident for $\beta_1$ and $\gamma_N$. The statistics for the ratio $\sigma_{WSX}(p_{\mu})/\sigma_{W}(p_{\mu})$ are as follows

\begin{displaymath}
\begin{array}{l}
\sigma_{WSZ}(\alpha_H)/\sigma_{W}(\alpha_H) = 0.52 \ \ , \ \ (0.46, 0.57) \ \ , \ \  (0.44, 0.60) \ \ , \\
\sigma_{WSZ}(\beta_1)/\sigma_{W}(\beta_1) = 0.26 \ \ , \ \ (0.14, 0.55) \ \ , \ \ (0.11, 0.84) \ \ , \\
\sigma_{WSZ}(\gamma_N/\sigma_{W}(\gamma_N) = 0.18 \ \ , \ \ (0.12, 0.26) \ \ , \ \ (0.10, 0.33) \ \ , \\
\end{array}
\end{displaymath}
where $\sigma_{W}(p_{\mu})$ and $\sigma_{WSX}(p_\mu)$ are the errors from WL only and WL\,+\,SZ\,+\,XR data.

From now on, we will only discuss the constraints from WL\,+\,SZ\,+\,XR data starting from Fig.\,\ref{fig: fmgrref} which shows $\sigma(p_{\mu})$ for the reference case as a function of halo redshift, mass, and concentration. Although larger than the theoretical priors assumed in this work (i.e., $-0.25 \le \alpha_H \le 0.25$ and $-0.15 \le \beta_1 \le 0.15$), the constraints we get are nevertheless remarkable. They are definitely strong than the ones obtained in \cite{Salzano:2017qac} fitting the convergence and electron density data only for a subset of DHOST theories. Moreover, we are here able to constrain $\gamma_N$ too which is typically not included as a parameter being hold fixed by the requirement $G_{N}^{eff} = G_N$.

\begin{figure*}
\centering
\includegraphics[width=0.65\columnwidth]{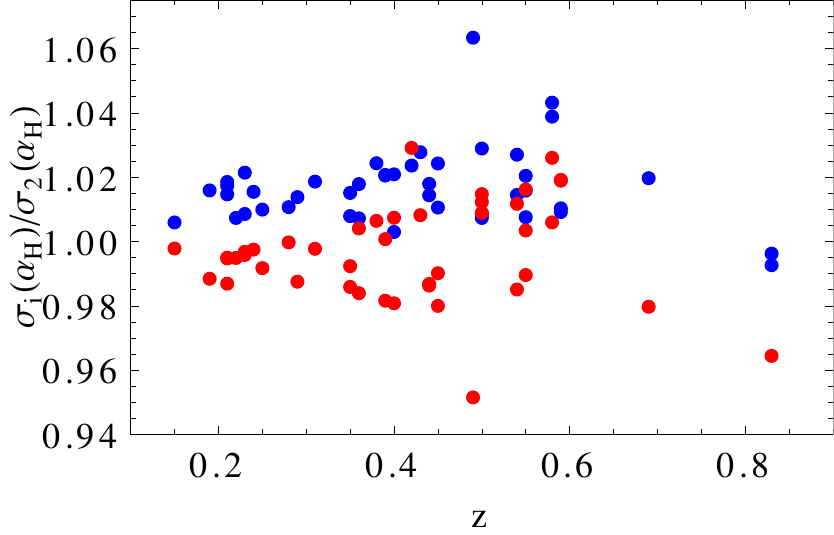}
\includegraphics[width=0.65\columnwidth]{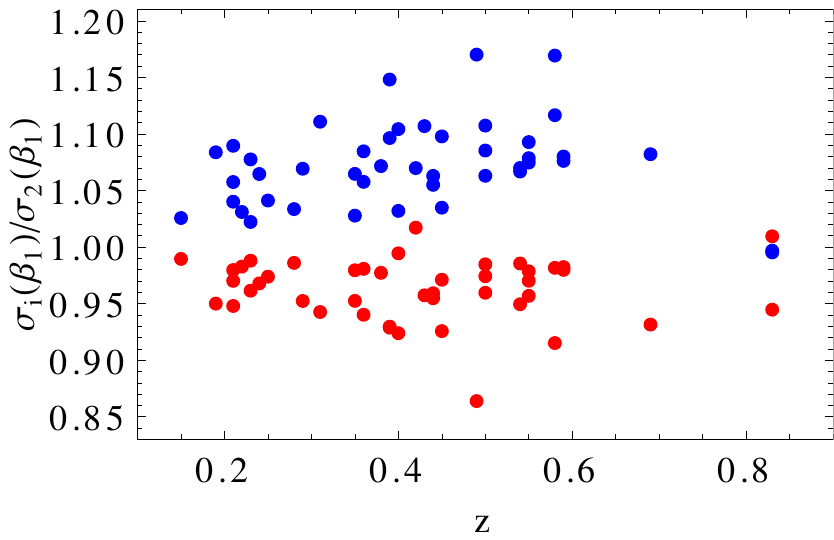}
\includegraphics[width=0.65\columnwidth]{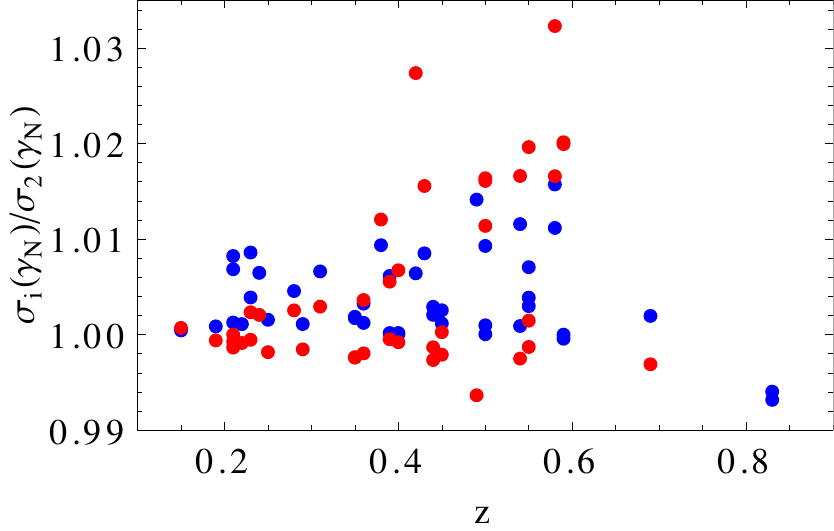} \\
\includegraphics[width=0.65\columnwidth]{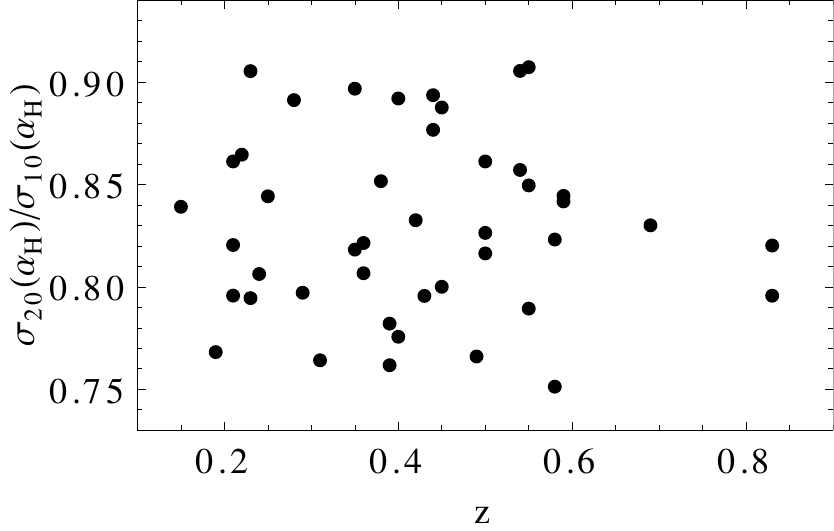}
\includegraphics[width=0.65\columnwidth]{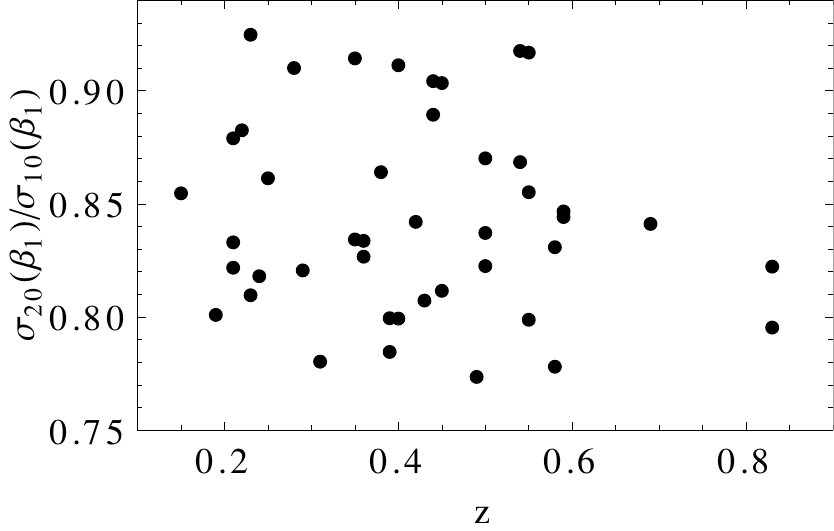}
\includegraphics[width=0.65\columnwidth]{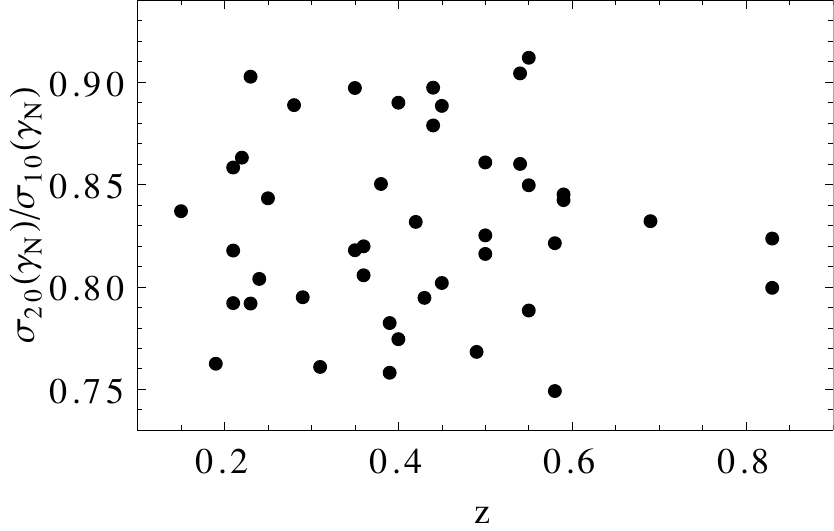} \\
\caption{Dependence of the constraints on the parameters $(\alpha_H, \beta_1, \gamma_N)$ on the SZ data radial range and the number of points in the WL and SZ dataset. {\it Top.} $\sigma_{i}(p_{\mu})/\sigma_2(p_{\mu})$ vs $z$ having denoted with $\sigma_{i}(p_\mu)$ the constraints from the {\it pessimistic}, {\it realistic}, {\it optimistic} case for $i = 1, 2, 3$ and used blue (red) points for $i = 1 (3)$. {\it Bottom.} $\sigma_{20}(p_{\mu})/\sigma_{10}(p_{\mu})$ vs $z$ being $\sigma_{n}(p_\mu)$ the constraints assuming $n$ points in the WL and SZ datasets.}
\label{fig: fmgrpesopt}
\end{figure*}

\begin{figure*}
\centering
\includegraphics[width=0.65\columnwidth]{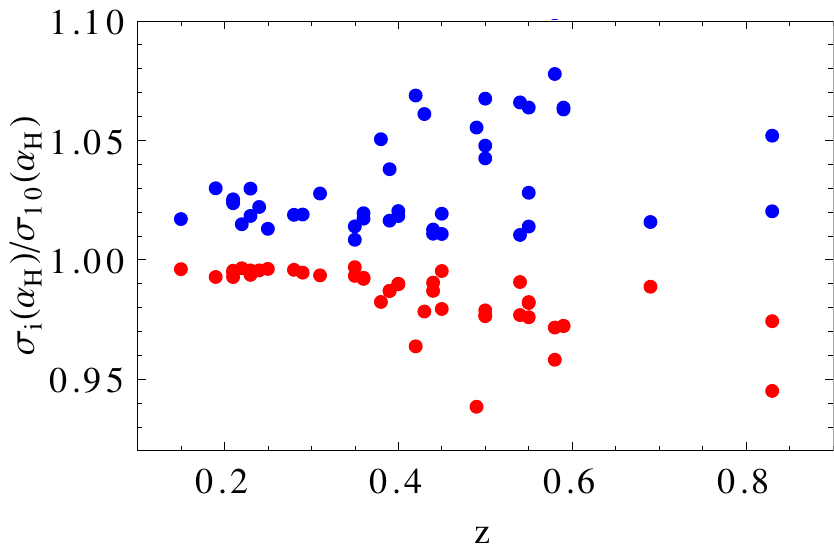}
\includegraphics[width=0.65\columnwidth]{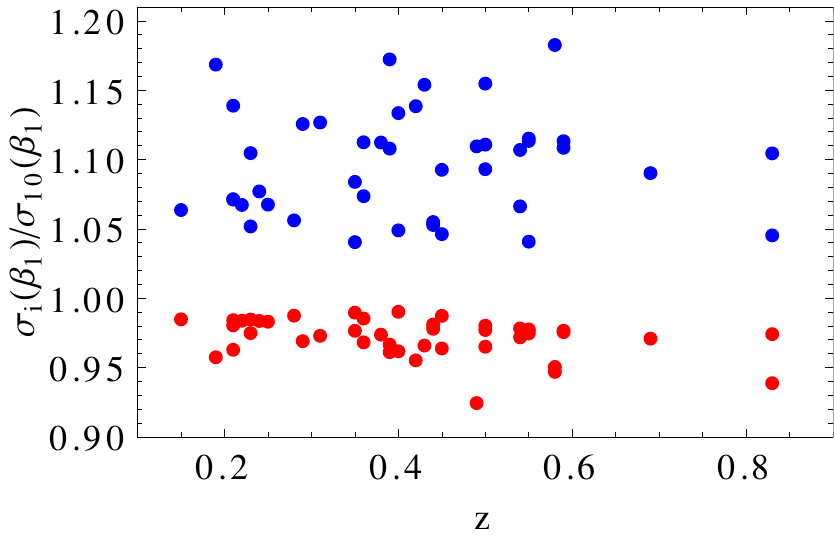}
\includegraphics[width=0.65\columnwidth]{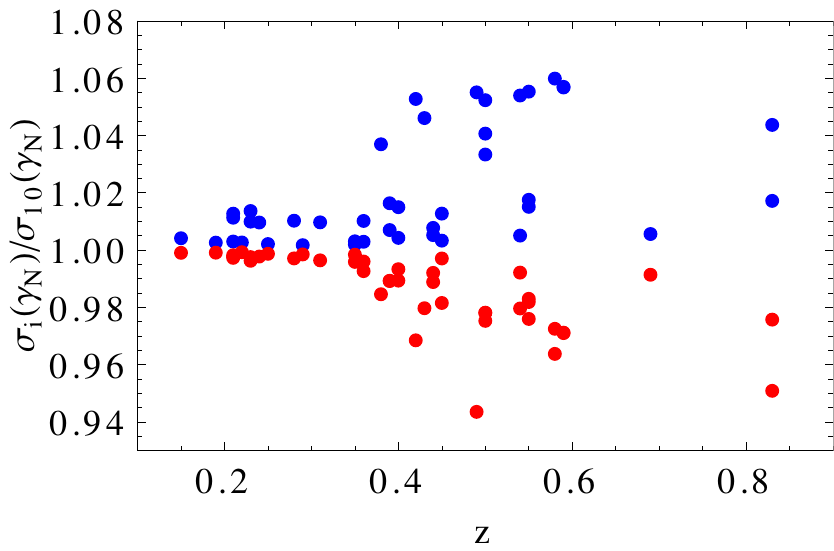} \\
\includegraphics[width=0.65\columnwidth]{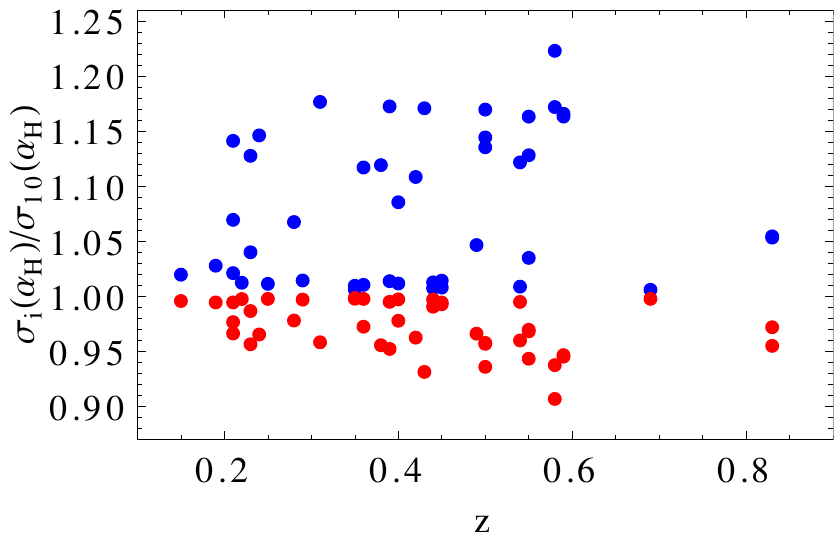}
\includegraphics[width=0.65\columnwidth]{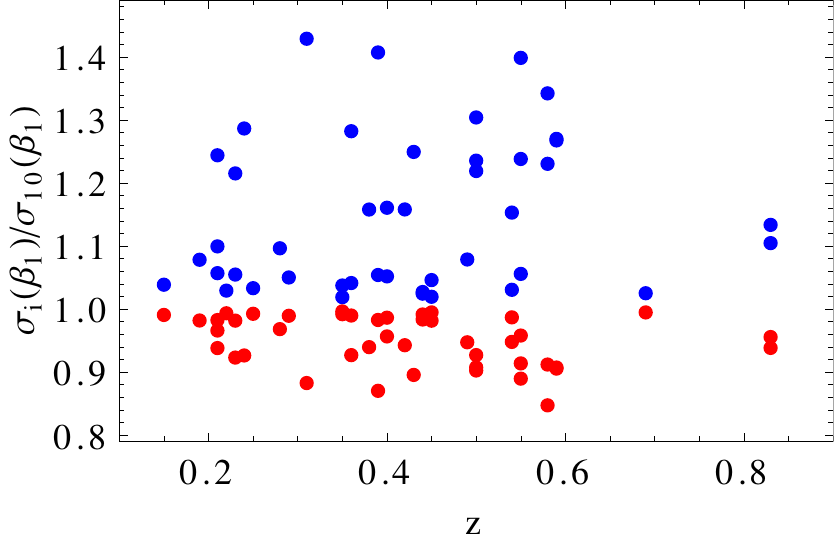}
\includegraphics[width=0.65\columnwidth]{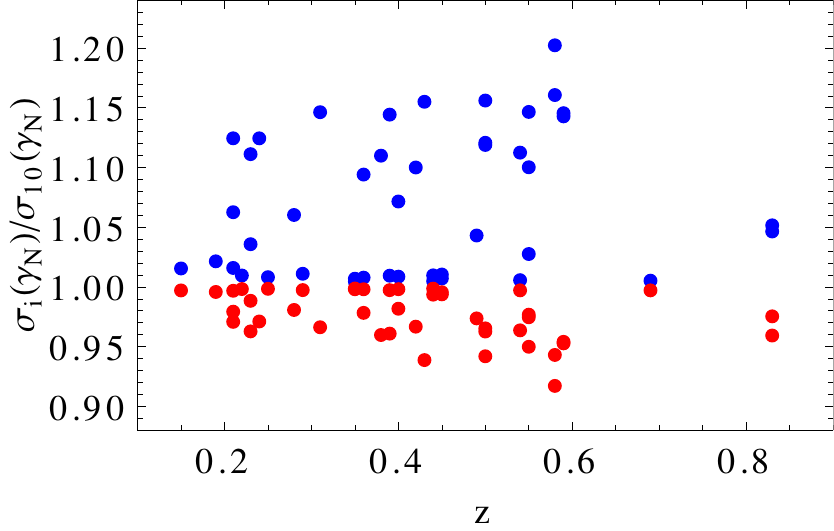} \\
\includegraphics[width=0.65\columnwidth]{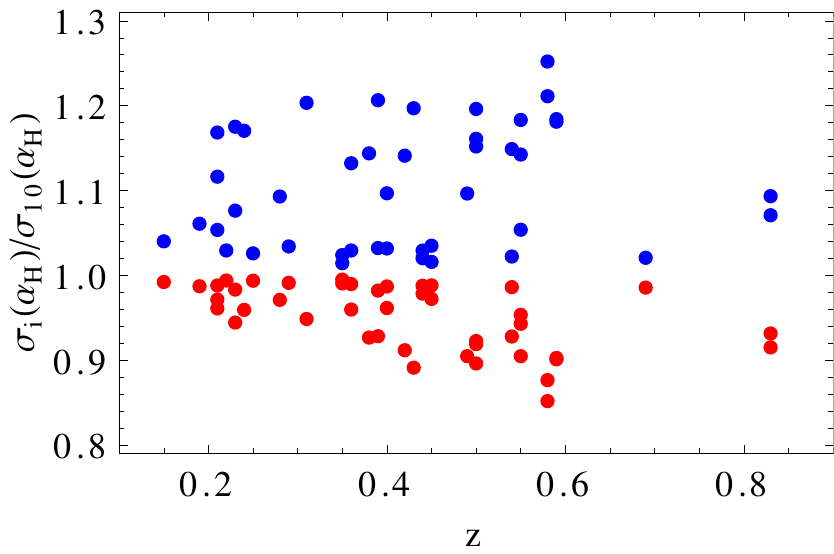}
\includegraphics[width=0.65\columnwidth]{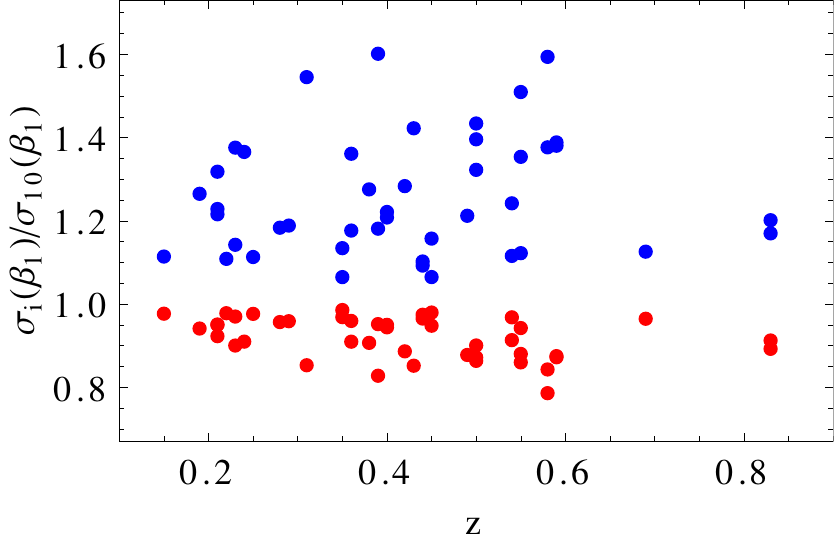}
\includegraphics[width=0.65\columnwidth]{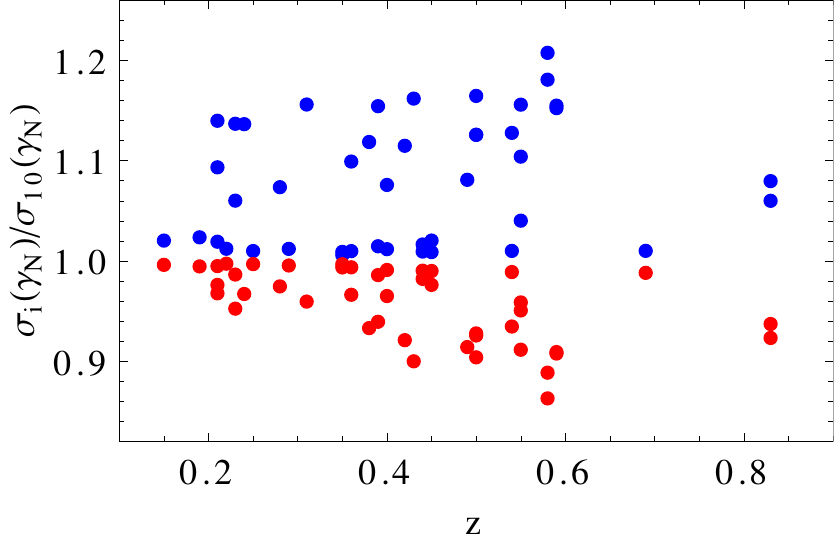} \\
\caption{Dependence of the constraints on the amplitude of the SZ and XR S/N scaling with distance and mass. {\it Top.} $\sigma_{b}(p_{\mu})/\sigma_{10}(p_{\mu})$ with $b = (05, 20)$ for ${\cal{B}}_e = (0.5, 1.5)$ for (blue, red) points keeping ${\cal{B}}_P = 1.0$, and all other setup quantities the same as for the reference case. {\it Centre.} Same as before but keeping ${\cal{B}}_e = 1$ and setting ${\cal{B}}_P = (0.5, 1.5)$ for (blue, red) points. {\it Bottom.} Same as before but now setting ${\cal{B}}_e = {\cal{B}}_P = (0.5, 1.5)$ for (blue, red) points.}
\label{fig: fmgrboost}
\end{figure*}

Fig.\,\ref{fig: fmgrref} shows a clear correlation between the errors and the cluster redshift pointing at high $z$ systems as most efficient target to constrain the DHOST parameters. This is likely related to our choice of holding fixed the angular range for the WL data. The larger is $z$, the more one is pushing data in the outer region $R > R_{200}$. Although the DHOST correction fades away with $R$, probing outer regions allows to better constrain the halo mass $\log{M_{200}}$ hence breaking both the degeneracy with $c_{200}$, and the one with $\gamma_N$. The effect is then propagated on the other parameters too thus qualitatively explaining the anticorrelation between $\sigma(p_{\mu})$ and $z$. Such a qualitative argument, however, should not be overrated since the observed trend with $z$ could also be a fake artifact of our choice of taking fixed the number of radial bins with $z$. Since clusters at higher $z$ have a smaller angular size, taking the number of bins fixed is possible only if we assume that the angular resolution of the instrument is enough to achieve this goal at all $z$. Whether this is the case or not depends on the observational setup, a point that we do not address in this forecasts analysis.

There is, on the contrary, no correlation with the halo mass and concentration. Although it is true that increasing more massive clusters have a better WL, SZ, and XR S/N, this is not sufficient to break any degeneracy among parameters hence not contributing to improve the constraints which explains the lack of correlation with $\log{M_{200}}$. On the other hand, the $c_{200}$ range probed by our systems is probably too small to find a signature of correlation with the concentration so that we invite the reader to not overrate the no correlation in the right panel of Fig.\ref{fig: fmgrref}. We also remind the reader that we have used a $c_{200}$\,-\,$M_{200}$ relation to set the fiducial value of the concentration of BOXSZ clusters neglecting its scatter. Should we have included it, the $c_{200}$ range would have been wider, but this would have asked to repeat the analysis for any realization of $c_{200}$ which is definitely too time consuming.

\begin{table*}
\begin{center}
\resizebox{\textwidth}{!}{
\begin{tabular}{cccccccccc} 
\hline 
 & \multicolumn{3}{c}{$\alpha_H$} & \multicolumn{3}{c}{$\beta_1$}
& \multicolumn{3}{c}{$\gamma_N$} \\
\hline
id & med & $68\%$ CL & $95\%$ CL & med & $68\%$ CL & $95\%$ CL  & med & $68\%$ CL & $95\%$ CL  \\
\hline \hline
GR & 1.351 & (1.148, 1.472) & (1.094, 1.682) & 0.218  & (0.181, 0.244)  & (0.171, 0.287) & 1.098 & (0.962, 1.227) & (0.917, 1.397)\\
\hline
A0M & 0.359 & (0.311, 0.444) & (0.278, 0.470) & 0.090 & (0.084, 0.106) & (0.077, 0.152) & 0.184 & (0.141, 0.223) & (0.120, 0.244) \\
A0P & 0.332 & (0.255, 0.435) & (0.222, 0.501) & 0.026 & (0.023, 0.031) & (0.020, 0.040) & 0.159 & (0.116, 0.220) & (0.097, 0.255) \\
\hline
B0M & 0.295 & (0.248, 0.395) & (0.202, 0.439) & 0.056 & (0.051, 0.060) & (0.047, 0.064)& 0.119 & (0.094, 0.153) & (0.078, 0.188) \\
B0P & 0.192 & (0.142, 0.261) & (0.127, 0.369) & 0.037 & (0.034, 0.044) & (0.030, 0.063) & 0.154 & (0.113, 0.235) & (0.097, 0.346) \\
\hline
D1M & 1.006 & (0.866, 1.204) & (0.845, 1.448) & 0.285 & (0.237, 0.336) & (0.227, 0.423) & 0.396 & (0.347, 0.475) & (0.328, 0.569) \\
D1P & 1.174 & (1.022, 1.347) & (1.003, 1.547) & 0.107 & (0.090, 0.121) & (0.086, 0.149) & 0.504 & (0.446, 0.585) & (0.438, 0.673) \\
\hline
D2M & 0.618 & (0.541, 0.702) & (0.478, 0.742) & 0.101 & (0.092, 0.119) & (0.086, 0.124) & 0.422 & (0.362, 0.483) & (0.320, 0.508) \\
D2P & 1.068 & (0.864, 1.195) & (0.765, 1.250) & 0.063 & (0.054, 0.078) & (0.052, 0.079) & 0.260 & (0.220, 0.305) & (0.191, 0.317) \\
\hline
\end{tabular}}
\caption{Median, $68\%$, and $95\%$ CL of $\sigma(p_{\mu})/|1 + p_{\mu}|$ with $p_{\mu} = (\alpha_H, \beta_1, \gamma_N)$ for the representative DHOST models listed in Table\,\ref{tab: dhostfid} assuming the reference configuration for the radial range, sampling, and S/N of the WL, SZ, and XR data.}
\label{tab: fmdhost}
\end{center}
\end{table*}

It is worth wondering how robust are the results with respect to the assumptions on the data. This would also allow to identify where efforts should be directed to improve the constraining power of the WL\,+\,SZ\,+\,XR datasets. To this end, we first look at how the constraints change when we change the radial range probed by SZ and XR data. Top panels in Fig.\,\ref{fig: fmgrpesopt} show the ratio of the constraints on $(\alpha_H. \beta_1, \gamma_N)$ parameters with respect to the reference case\footnote{We plot the ratio as function of the redshift just to better separate objects in the plot, but what matters here are just the numbers on the $y$\,-\,axis.}  when we move to the {\it central} and {\it large range} scenario. The most affected parameter is $\beta_1$ with the error increasing (decreasing) up to $\sim 20\%$ $(\sim 15\%)$ when reducing (increasing) $\xi_{max}$ to 0.87 (1.81) from the fiducial $\xi_{max} = 1.17$ case. This is likely related to the results in Fig.\,\ref{fig: prvsab} showing that the deviations of $P(r)$ from its GR counterpart are larger for larger $\xi = r/R_{500}$. However, it is worth noting that the improvement obtained by increasing the radial range probed by the data is actually not so large with the typical decrease of $\sigma(\beta_1)$ being of order $5\%$ at the cost of a $55\%$ increase of $\xi_{max}$. We therefore argue that this is not the most convenient way to improve the constraints. On the contrary, increasing the angular resolution, i.e., using 20 instead of 10 points for both WL and SZ data, has a major impact on the constraints. As bottom panels in Fig.\,\ref{fig: fmgrpesopt} show, the error improves by $\sim 15 - 20\%$ for all the DHOST parameters as a consequence of the larger number of terms in the sum in Eqs.(\ref{eq: kappafmbis}) and (\ref{eq: pressfm}). Note that this result could not have been easily anticipated since increasing ${\cal{N}}_{WL}$ reduces the WL S/N because of the smaller ${\cal{A}}$ in Eq.(\ref{eq: stonkappa}). We nevertheless find that the dominating effect is the increased number of terms in the sum defining the Fisher matrix elements. We therefore recommend a finer sampling as a way to improve the constraints on DHOST parameters.

Finally, we have also investigated how the constraints depend on our assumption on the data S/N by varying $({\cal{B}}_e, {\cal{B}}_P)$ which scale up or down the amplitude of $\nu_e(r/R_{500}, \log{M_{500}})$ and $\nu_P(r/R_{500}, \log{M_{500}})$, respectively. The results are shown in Fig.\,\ref{fig: fmgrboost} where all the ratios are taken with respect to the reference case, and we vary only $({\cal{B}}_e, {\cal{B}}_P)$ taking the radial range and sampling unaltered. As a first case, we show in top panels the impact of ${\cal{B}}_e$ degrading or boosting the electron density S/N by $50\%$ (i.e., ${\cal{B}}_e = 0.5$ or ${\cal{B}}_e = 1.5$). The parameter most affected by a degradation of the XR S/N is again $\beta_1$. In order to understand why this happens, we notice that $\beta_1$ enters the correction to the convergence only through the third term in Eq.(\ref{eq: calsdef}) which is subdominant compared to the first two. As a consequence, $\beta_1$ is mainly constrained by the pressure data. A degradation of the S/N of XR data worsens the constraints on the double\,-\,$\beta$ profile parameters hence weakening the constraining power of the pressure data too. On the contrary, both $\alpha_H$ and $\gamma_N$ are constrained by both WL and SZ so that are less affected by a weakening of the constraints from electron density. This same argument also explains the results shown in the second row panels of Fig.\,\ref{fig: fmgrboost} where we set back ${\cal{B}}_e = 1$, but investigate the impact of varying ${\cal{B}}_P$ to 0.5 or 1.5 from the fiducial value. However, a degradation of the pressure ${\rm S/N}$ impacts all the parameters since it makes harder to break the degeneracy present if one uses WL data alone. Needless to say, degrading by $50\%$ both XR and SZ data has a dramatic effect on the constraints with the errors which increase by up to $\sim (27, 60, 20)\%$ on $(\alpha_H, \beta_1, \gamma_N)$. It is worth stressing, however, that such a severe overestimate of the S/N is quite unlikely. Indeed, we have derived our S/N scaling functions based on up to date dataset spanning a similar mass range as the BOXSZ cluster sample. The difference in redshit is not critical, and the instrumental setup is well representative of what can be achieved with present facilities. On the other hand, it is also worth noticing that a sort of saturation takes place when increasing the S/N. Indeed, a $50\%$ boost of both the SZ and XR S/N (corresponding to the red points in the bottom panels of Fig.\,\ref{fig: fmgrboost}) causes a reduction of the errors by a modest $\sim (5, 10, 5)\%$ for $(\alpha_H, \beta_1, \gamma_N)$. This is likely related to the choice of taking unaltered the WL S/N since it is this dataset to play the major role. As a concluding remark, we can summarise the results of the analysis of variation of the constraints with the different observational quantities in a single take\,-\,home lesson. In order to improve the constraints on the DHOST parameters, the better strategy is to use the present setup in terms of S/N and radial range probed, but increase the sampling of the convergence and pressure profiles.

\begin{table*}
\begin{center}
\begin{tabular}{ccccc} 
\hline 
\hline
id & bin no. 1 & bin no. 2 & bin no. 3 & bin no. 4 \\
\hline \hline
GR & (0.437, 0.072, 0.364) & (0.380, 0.059, 0.318) & (0.365, 0.058, 0.302) &  (0.335, 0.053 ,0.278) \\
\hline
A0M & (0.100, 0.026, 0.050) & (0.105, 0.025, 0.053) & (0.093, 0.025, 0.044) &  (0.092, 0.025, 0.045) \\
A0P & (0.083, 0.007, 0.041) & (0.091, 0.007, 0.045) & (0.077, 0.007, 0.036) & (0.076, 0.007, 0.036) \\
\hline
B0M & (0.072, 0.016, 0.029) & (0.085, 0.015, 0.0345) & (0.081, 0.015, 0.031) & (0.078, 0.015, 0.030) \\
B0P & (0.043, 0.011, 0.037) & (0.050, 0.011, 0.042) & (0.046, 0.010, 0.037) & (0.045, 0.010, 0.036) \\
\hline
D1M & (0.344, 0.096, 0.136) & (0.288, 0.076, 0.115) & (0.283, 0.076, 0.112) & (0.259, 0.069, 0.102) \\
D1P & (0.382, 0.035, 0.167) & (0.334, 0.029, 0.147) & (0.330, 0.029, 0.144) &  (0.305, 0.027, 0.133) \\
\hline
D2M & (0.165, 0.028, 0.113) & (0.188, 0.030, 0.128) & (0.177, 0.029, 0.120) & (0.165, 0.027, 0.111) \\
D2P & (0.284, 0.018, 0.072) & (0.313, 0.019, 0.080) & (0.297, 0.018, 0.075) & (0.270, 0.016, 0.068) \\
\hline
\end{tabular} 
\caption{Relative errors $\sigma(p_{\mu})/|1 + p_{\mu}|$ for $p_{\mu} = (\alpha_H, \beta_1, \gamma_N)$ joining together the constraints from all the clusters in the same redshift bin. We consider all the DHOST models listed in Table\,\ref{tab: dhostfid} and set the observational setup as for the reference case. Note that, different from Table\,\ref{tab: fmdhost}, the values here refer to the error as estimated by the joint fit to all the clusters in a single bin rather than the median and $68\%$\,CL of the distribution of the errors from the fit to each single cluster.}
\label{tab: fmjoint}
\end{center}
\end{table*}

\subsection{Varying the DHOST fiducial parameters}

The above results rely on the assumption that the true underlying model is GR so that they tell us how well the data constrain deviations from GR itself. On the other hand, it is also interesting to investigate whether DHOST theories can be discriminated based on the values of their parameters. We therefore here look at the relative constraints, i.e., we compute $\sigma(|1 + \alpha_H|)/|1 + \alpha_H|$, $\sigma(|1 + \beta_1|)/|1 + \beta_1|$, and $\sigma(\gamma_N)/\gamma_N$ for the eight representative models listed in Table\,\ref{tab: dhostfid}. Note that we use absolute values and add unity to avoid any divergence of the ratios. We have computed the constraints for all the configurations discussed in the previous paragraph, but we are here interested only in investigating how they change depending on the fiducial DHOST parameters. This point turns out to be qualitatively unaffected by the particular choice of the radial range, the sampling, and the S/N amplitude so that we report in Table\,\ref{tab: fmdhost} only the constraints for the reference configuration.

As a general result, we find that the constraints for all models other than GR are stronger than for the GR itself. This is a naive consequence of the fact that it is easier to spot a signature when it is present in the data. Indeed, for all the DHOST fiducials, at least one of the quantities $(\Xi_1, \Xi_2, \Xi_3)$ is non vanishing, while all of them are zero for GR. As a consequence, data can constrain the amplitude of the corresponding non GR terms in the convergence and/or pressure profiles which makes it easier to constrain $(\alpha_H, \beta_1, \gamma_N)$. 

This same consideration also helps to understand why the worst constraints (apart from GR) are obtained for the D1M and D1P cases. Indeed, for these models, it is $\Xi_1 = \Xi_3 = 0$ so that the only deviation from GR is due to the second term in Eq.(\ref{eq: calsdef}) so that only the convergence can detect the signature of the corresponding DHOST models. A similar argument also applies for the D2M and D2P fiducials. In these cases, however, all the three amplitude parameters $(\Xi_1, \Xi_2, \Xi_3)$ are non vanishing, but $\Xi_2$ and $\Xi_3$ have opposite signs so that the two DHOST terms almost cancel each other. The DHOST signal is then present mainly in the pressure profile which is less constraining than convergence. As an overall consequence, the constraints on $(\alpha_H, \beta_1, \gamma_N)$ are weakened as evident from the larger $\sigma(p_{\mu})/|1 + p_{\mu}|$ values.

On the contrary, the strongest constraints are obtained when one of the two parameters $(\alpha_H, \beta_1)$ is set to zero. This is apparently in contradiction with what we have said about the GR case. Actually, this is not since one has to look at the values of $(\Xi_1, \Xi_2, \Xi_3)$ rather than to $(\alpha_H, \beta_1)$. In particular, for models A0M and A0P, one gets $\Xi_2 = 0$ so that the deviations from GR in the convergence profile are maximized since there is no more a compensation of the second and third term in Eq.(\ref{eq: calsdef}). A similar argument also applies to models B0M and B0P where it is now $\Xi_3 = 0$. The fact that constraints are stronger for B0M and B0P rather than A0M and A0P is finally related to the larger (absolute) value of $\Xi_1$ which makes it easier to detect the DHOST signature in the pressure profile.

\subsection{Constraints from joint use of more clusters}

As already said, the DHOST parameters $(\alpha_H, \beta_1, \gamma_N)$ are actually functions of the redshift so that two clusters with the same mass and concentration, but different $z$ experience a different deviation from GR. As a consequence, one can not stack clusters with the same $\log{M_{200}}$ to increase the lensing or the pressure signal hence the S/N and then decrease the error on the parameters of interest. However, should $(\alpha_H, \beta_1, \gamma_N)$ be slowly varying functions of $z$, one can combine the constraints from all the clusters in a given redshift bin. From the point of view of Fisher matrix forecasts, this is obtained by first marginalizing over all parameters other than the DHOST ones and then summing up the marginalized Fisher matrices. Note that this is not the same as stacking clusters in the same redshift bin and perform a fit to the stacked data. On the contrary, one is still fitting each single cluster data, but then multiply the different likelihood functions after marginalizing over all the astrophysical parameters.

To this end, we split the sample in 4 equipopulated redshift bins\footnote{Note that, in order to have the same number of clusters in each bin, we have adjusted the bin widths. The bin limits turn out to be $(0.15, 0.28)$, $(0.29, 0.40)$, $(0.42, 0.54)$, $(0.55, 0.83)$. Apart from the last one, all the bins have comparable width due to the almost uniform redshift distribution of the BOXSZ sample.} with median redshifts $(0.22, 0.36, 0.45, 0.58)$, each bin containing 11 cluster. We then compute the constraints from the combined marginalized Fisher matrices assuming the reference configuration for radial range, sampling, and S/N. The results thus obtained are summarized in Table\,\ref{tab: fmjoint} where we consider all the models discussed before for the fiducial DHOST parameters. 

Comparing the median values in Table\,\ref{tab: fmdhost} wit the constraints in Table\,\ref{tab: fmjoint}, it is evident how joining clusters dramatically improves the contraints. The errors on all DHOST parameters are reduced by a factor $\sim 3$ in the GR case, while the improvement may differ depending on the particular DHOST models chosen. In all cases, it is, however, roughly comparable with ${\cal{N}}_{c}^{1/2}$, ${\cal{N}}_c$ being the number of clusters in the bin. Increasing ${\cal{N}}_{c}$ also helps in reducing the width of the bins thus reducing the systematic error related to the assumption that $(\alpha_H, \beta_1, \gamma_N)$ are constant within each redshift bin. 

Table\,\ref{tab: fmjoint} does not show a marked trend of the errors decreasing with the median redshift of the bin which would have been expected based on what is shown in Fig.\,\ref{fig: fmgrref}. This is likely a consequence of having mixed systems with different $z$ so that the trend is smoothed out\footnote{This result could be surprising at first sight since taking the median of the errors from the fit to the single clusters in each redshift bin gives a trend with $z$. However, the constraints from the joint fit we are considering here are not the median of the constraints from single objects.}. However, it is worth noticing that a complete analysis of the trend of $\sigma(p_{\mu})$ with $z$ should ask for the solution of the cosmological equation to get the values of $(\alpha_H, \beta_1, \gamma_N)$ to be used as fiducial for the Fisher matrix at each given $z$. As we have said, indeed, the constraints get weaker as the fiducial approaches GR. Since this is expected to happen as $z$ increases, two contrasting effects are at work. On one hand, going to larger $z$ with a fixed angular range for the WL data helps to improve the constraints since one is probing more and more into the region $R > R_{200}$. On the other hand, the fiducial approaches GR so that the DHOST corrections become smaller which weaken the constraints. Which effect is dominant depends on the specifics of the DHOST model, but a case\,-\,by\,-\,case analysis is outside our aims. 


\section{Conclusions}

Hunting for the responsible of the current cosmic speed up is one of the most fascinating and yet hardest challenge of present day cosmology. The landscape to be searched for is made wider by the consideration that the accelerated expansion may be taken as a first evidence for the need of a more general theory of gravitation. DHOST theories are ideal candidates from this point of view being able to produce an accelerated expansion of the universe without violating the constraints from GW and the stability and ghost\,-\,free requirements. A Vainshtein\,-\,like screening prevents fifth\,-\,force manifestation on Solar System scale, but modification to the gravitational potentials are still possible on galaxy clusters scale. It is therefore possible to look for signatures of DHOST theories in clusters observable quantities such as the lensing convergence $\kappa(R)$ and the pressure profile $P(r)$. We have here derived theoretical expressions for both of them and investigated the impact of the additional DHOST contributions. As a side result, we have also shown that the theoretical pressure profile may be well fitted by the empirical universal pressure profile with parameters which cover the same regions in the parameter space occupied by a subsample of Planck clusters. 

It turns out that the deviations from GR depend on the DHOST parameters in a different way for $\kappa(R)$ and $P(r)$ so that jointly fitting both quantities can help constraint DHOST parameters $(\alpha_H, \beta_1, \gamma_N)$ breaking degeneracy among them. The addition of X\,-\,ray data on the electron density do not constrain the DHOST parameters themselves, but gives an indirect yet fundamental contribution by constraining the electron density hence the astrophysical parameters determining the shape of the pressure profile. A Fisher matrix analysis has then been carried out in order to quantify how strong these constraints are under various assumptions on the observational setup. In particular, we have both investigated whether observations can discriminate between GR and DHOST, and how well $(\alpha_H, \beta_1, \gamma_N)$ may be constrained for some representative choices of their fiducial values. 

When using individual clusters and setting the fiducial model to GR, the constraints on $(\alpha_H, \beta_1, \gamma_N)$ are definitely improved by the joint use of the three probes with the errors reducing by factor $\sim (2, 4, 5)$ with respect to those from WL alone. The $68\%$\,CL are, however, still larger than the theoretical priors one can obtain by asking that the background evolution is not too different from that for the concordance $\Lambda$CDM model and that there are no ghosts or instabilities. This is not surprising since the theoretical requirements apply to the full evolutionary history, while our results refer to the value of the parameters at the given cluster redshift. However, the approach we present here is not biased by any theoretical prejudice since it only relies on the comparison with data. It should therefore be considered as complementary to a theoretical analysis. On the other hand, one could also try to combine the two methods adding a prior on $(\alpha_H. \beta_1, \gamma_N)$ informed by the theory requirements. However, we argue against such a possibility since it makes the likelihood far from Gaussian which is contrary to the assumption underlying the Fisher matrix methodology. One should rather perform a fit to mock data including the theory requirements as hard priors in the MCMC sampling of the parameters space, a possibility that we will explore in a future work.

A similar discussion qualitatively hold also when one changes the fiducial values of $(\alpha_H, \beta_1, \gamma_N)$ as we have shown considering eight representative DHOST models. The constraints, however, result to be stronger than in the GR case the strengthening being determined by the values of $(\Xi_1, \Xi_2, \Xi_3)$ for the given model. This is a naive consequence of the fact that these parameters set the amplitude of the terms which make the convergence and pressure profile for DHOST theories deviate from their GR counterparts. For all models, however, what is of great help is the joint analysis of more clusters in the same redshift bin under the assumption that $(\alpha_H, \beta_1, \gamma_N)$ are approximately constant over the bin redshift range. This joint analysis, indeed, improves the constraints by a factor $\sim {\cal{N}}_{c}^{1/2}$ with respect to the median error from individual clusters. One should therefore aim at assembling a large sample of clusters to be split in narrow redshift bins in order to both increase the accuracy on the parameters and make the assumption of them being constant much more solid. Supposing we want to probe the evolution of $(\alpha_H, \beta_1, \gamma_N)$ over the range $z_{min} \le z \le z_{max}$, the total number of clusters should be ${\cal{N}}_{tot} \simeq (\Delta z/\delta z) f^2$ with $\Delta z = z_{max} - z_{min}$, $\delta z$ the bin width, and $f$ by how much we want to improve the constraints. For $(z_{min}, z_{max}) = (0.1, 0.9)$, $\delta z = 0.05$, $f = 5$, we get ${\cal{N}}_{tot} = 400$, a factor 9 increase with respect to the BOXSZ sample. 

It is worth wondering how such a dataset should be assembled, i.e., whether one must invest efforts in observing low or high redshift objects, select them according to the mass, investing time resources to improve the S/N or the sampling. We find that the errors on $(\alpha_H. \beta_1, \gamma_N)$ anti correlate with the cluster redshift, i.e., the higher $z$, the smaller $\sigma(p_{\mu})$. Such a result is related to our choice of measuring the convergence over a fixed angular range $(0.2 \le \theta/{\rm arcmin} \le 10.2)$ which makes the constraint on $\log{M_{200}}$ for high $z$ clusters better since one is pushing the data in the $R > R_{200}$ region. A reanalysis is, however, needed to take into account the variation of $(\alpha_H, \beta_1 \gamma_N)$ with $z$ in order to check whether the fact that any DHOST theory should reduce to GR at high $z$ does not degrade the constraints from objects in higher redshift bins. 

Somewhat unexpectedly, we find that our reference configuration with the electron density sampled with $40$ linearly spaced points over the range $(0.1, 1.17) R_{500}$ and ${\cal{N}}_{WL} = {\cal{N}}_{SZ} = 10$ points to sample the convergence and pressure profiles is nearly optimal. There is only a relatively modest reduction of the errors on DHOST parameters if one tries to push the upper limit of the pressure data to $1.81 R_{500}$ or to boost the XR and SZ S/N by a factor 1.5 with respect to the fiducial values we have assumed. The most efficient way to reduce the errors turns out to be a better angular resolution, i.e. doubling ${\cal{N}}_{WL}$ and ${\cal{N}}_{SZ}$. Although a mismatch in the estimate of the way S/N scales with distance and halo mass is possible, we are confident that this result is quite robust. Indeed, we have estimated the WL S/N based on what is expected for a Euclid\,-\,like setup, while the XR and SZ S/N have been estimated from real data. We therefore plan to investigate the feasibility of such a hypothetical survey based on simulated mock data to strengthen this preliminary result. A word of caution concerns, in particular, the possibility to have ${\cal{N}}_{WL} = {\cal{N}}_{SZ}$ for all the clusters in the sample no matter their redshift. While having ${\cal{N}}_{WL} = 10$ or $20$ is quite easy to achieve, some more efforts are needed to get the same value for ${\cal{N}}_{SZ}$. Assuming to have an angular resolution of $\sim 20 \ {\rm arcsec}$ over the range $(0.1, 1.2) R_{500}$ and using the estimated values of the mass and redshift for the BOXSZ clusters, we get ${\cal{N}}_{SZ} = 14$ as median value with trend in redshift from ${\cal{N}}_{SZ} = 30$ at $z = 0.15$ to ${\cal{N}}_{SZ} = 7$ at $z = 0.83$ (with a scatter due to the dependence on mass). Our choice ${\cal{N}}_{SZ} = 10$ is therefore in between the extreme cases, while getting ${\cal{N}}_{SZ} = 20$ can indeed be harder for the higher $z$ clusters unless they are massive enough to guarantee a large $R_{500}$. Investigating how the constraints change depending on the median and the scatter of the ${\cal{N}}_{SZ}$ distribution is outside our aims here, but it is a point not to be forgotten in an analysis based on mock data.

As a final remark, we want to come back to our adopted strategy to use cluster data. Here, we have adopted the backward approach using an empirical profile to fit the electron density, the NFW model for the dark halo convergence, and the solution of the hydrostatic equilibrium equation for the pressure profile. Although vastly used in the literature, this method relies on a number of reasonable yet aprioristic assumptions. A more empirically based alternative is, actually, possible. One could, indeed, choose phenomenological models for both the gas density and the pressure profile (e.g., the Vikhlinin and the universal profile, respectively), fit them to the X\,-\,ray and SZ data, and then plug the results into the hydrostatic equilibrium equation to derive a theoretical dark halo mass profile. This could be later compared to the lensing convergence data thus constraining both the cluster and DHOST parameters. A forthcoming companion paper will present the results of this method. We nevertheless anticipate that choosing among them is more a matter of how much one trust the underlying assumptions of each method. A safer option when dealing with real data would therefore be to use both of them and compare the results as a consistency test.

Being the largest bound structures in the universe, galaxy clusters have always been looked at as ideal laboratories for testing the theory of gravitation. The present work confirms that this is indeed the case for DHOST models too.

\section*{Acknowledgments}

All the authors acknowledge support from INFN/Euclid Sezione di Roma. PK, MDP and RM also acknowledge support from Sapienza Universit\'a di Roma thanks to Progetti di Ricerca Medi 2018, prot. RM118164365E40D9 and 2019, prot. RM11916B7540DD8D.


\appendix


\section{DHOST action}\label{app:dhost}

The GW170817 event and the tiny delay in the arrival of signal from its electromagnetic counterpart has severely restricted the class of viable DHOST theories. For the surviving models, the action can be written as

\begin{equation}
S = \int{d^4x \sqrt{-g} {\cal{L}}}
\label{ac: total}
\end{equation}
where the integrand Lagrangian is given by

\begin{eqnarray}
{\cal{L}} & = & P + Q \Box \phi + F R + A_3 \phi^{\mu} \phi^{\nu} \phi_{\mu \nu} \Box \phi \nonumber \\
 & + & \frac{48 F_{X}^{2} - 8 (F - X F_X) A_3 - X^2 A_{3}^{2}}{8F} \ \phi^{\mu} \phi_{\mu \nu} \phi_{\lambda} \phi^{\lambda \nu}
\nonumber \\
 & + & \frac{(4 F_X + X A_3) A_3}{2F} \ (\phi_{\mu} \phi^{\mu \nu} \phi_{\nu})^2
\label{ac:dhost:gw}
\end{eqnarray}
with $(P, Q, F, A_3)$ arbitrary functions of the scalar field $\phi$ and its kinetic energy $X$, and the label denoting derivative with respect to $X$. GR is recovered setting $F = 1/2\kappa$ with $\kappa = 8\pi G/c^{4}$, and $P = Q= A_3 = 0$. In the general case, one can adopt the EFT formalism \cite{Langlois:2017mxy} to express the functions $(P, Q, F, A_3)$ in terms of the time dependent linear operators $(\alpha_T, \alpha_M, \alpha_K, \alpha_H, \beta_1)$. The constraints from the GW170817 event forces to set $\alpha_T = 0$, while only $(\alpha_H, \beta_1)$ enter the weak field limit causing the deviations of $(\Phi, \Psi)$ potentials from their GR counterparts as shown in Eqs.(\ref{eq: gravforceone})\,-\,(\ref{eq: gravforcetwo}).


\section{Derivatives for Fisher matrix computation}\label{app:fmder}

We report below all the derivatives needed to compute the convergence and pressure profile Fisher matrices. As a general rule, all these derivatives ask for numerical integration. We have, however, checked that all the integrals are quite stable so there are no numerical issues affecting the estimate of the Fisher matrix.

\subsection{Lensing convergence}

The parameters the lensing convergence depends are $\{\log{M_{200}}, c_{200}, \gamma_N, \alpha_H, \beta_1\}$. The relevant derivatives are quite straightforward to compute giving

\begin{equation}
\frac{\partial \ln{\kappa}(R)}{\partial \log{M_{200}}} = 
\frac{\ln{10}}{3} \left [ 1 - 
\left ( \frac{R}{R_v} \right )^2 
\frac{{\cal{K}}_{\xi}(R/R_v; c_{200}; \alpha_H, \beta_1)}
{{\cal{K}}_0(R/R_v; c_{200}; \alpha_H, \beta_1)}
\right ] \ ,  
\label{eq: dlnkdlogmv}
\end{equation}

\begin{eqnarray}
\frac{\partial \ln{\kappa}(R)}{\partial c_{200}} & = & 
\frac{2}{c_{200}} - \frac{c_{200}/(1 + c_{200})^2}{\ln{(1 + c_{200})} - c_{200}/(1 + c_{200})} \nonumber \\
 & \times & 
\frac{{\cal{K}}_c(R/R_v; c_{200}; \alpha_H, \beta_1)}
{{\cal{K}}_0(R/R_v; c_{200}; \alpha_H, \beta_1)} \ , 
\label{eq: dlnkdcv}
\end{eqnarray}

\begin{equation}
\frac{\partial \ln{\kappa}(R)}{\partial \gamma_N} = \frac{1}{\gamma_N} \ ,
\label{eq: dlnkdgammaN}
\end{equation}

\begin{equation}
\frac{\partial \ln{\kappa}(R)}{\partial \alpha_H} = 
\frac{1}{1 - \alpha_H - 3 \beta_1} +
\frac{{\cal{K}}_{\alpha}(R/R_v; c_{200}; \alpha_H, \beta_1)}
{{\cal{K}}_0(R/R_v; c_{200}; \alpha_H, \beta_1)} \ , 
\label{eq: dlnkdalphaH}
\end{equation}

\begin{equation}
\frac{\partial \ln{\kappa}(R)}{\partial \alpha_H} = 
\frac{3}{1 - \alpha_H - 3 \beta_1} +
\frac{{\cal{K}}_{\beta}(R/R_v; c_{200}; \alpha_H, \beta_1)}
{{\cal{K}}_0(R/R_v; c_{200}; \alpha_H, \beta_1)} \ .
\label{eq: dlnkdbeta1}
\end{equation}
Here we have defined the ${\cal{K}}_p$ functions above as 

\begin{equation}
{\cal{K}}_{p}(R/R_v; c_{200}; \alpha_H, \beta_1) = 
\int_{-\infty}^{\infty}{{\cal{S}}_p(y; c_{200}; \alpha_H, \beta_1) d\zeta}
\label{eq; ktildefundef}
\end{equation}
with $y = (\xi^2 + \zeta^2)^{1/2} = (R^2/R_v^2 + z^2/R_v^2)^{1/2}$, and 

\begin{equation}
\begin{array}{l}
\displaystyle{{\cal{S}}_0(y; c_{200}; \alpha_H, \beta_1) = } \\ 
 \\ 
\displaystyle{\frac{2(1 - \beta_1) + (3 \alpha_H + \beta_1 + 4) c_{200} y + 
2 c_{200}^2 y^2}{2 y (1 + c_{200} y)^4}} \ ,
\end{array}
\label{eq: defS0}
\end{equation}

\begin{equation}
\begin{array}{l}
\displaystyle{{\cal{S}}_{\xi}(y; c_{200}; \alpha_H, \beta_1) =} \\
 \\ 
\displaystyle{\frac{(\beta_1 - 1)(1 + 5 c_{200} y) - (6 \alpha_H + 2 \beta_1 + 7) c_{200}^2 y^2 - 3 c_{200}^3 y^3}{y^3 (1 + c_{200} y)^5}} \ ,
\end{array}
\label{eq: defSxi}
\end{equation}

\begin{equation}
\begin{array}{l}
\displaystyle{{\cal{S}}_{c}(y; c_{200}; \alpha_H, \beta_1) =} \\
 \\ 
\displaystyle{\frac{(3 \alpha_H + 9 \beta_1 - 4) - (9 \alpha_H + 3 \beta_1 + 8) c_{200} y - 4 c_{200}^2 y^2}{2 (1 + c_{200} y)^5}} \ ,
\end{array}
\label{eq: defSc}
\end{equation}

\begin{equation}
{\cal{S}}_{\alpha}(y; c_{200}; \alpha_H, \beta_1) = 
\frac{3 c_{200}}{2 (1 + c_{200} y)^4} \ ,
\label{eq: defSa}
\end{equation}

\begin{equation}
{\cal{S}}_{\beta}(y; c_{200}; \alpha_H, \beta_1) = 
\frac{c_{200} y - 2}{2 y (1 + c_{200} y)^4} \ . 
\label{eq: defSb}
\end{equation}

\subsection{Pressure profile}

Eq.(\ref{eq: prdhost}) shows that the pressure profile depends on a large number of parameters. These are needed to fix the cluster properties $(\log{M_{200}}, c_{200})$, the electron density double\,-\,$\beta$ profile $(\beta, \log{x_{c1}}, \log{x_{c2}}, \log{n_{01}}, \log{(n_{02}/n_{01})})$, and the DHOST quantities $(\alpha_H, \beta_1, \gamma_N)$. The derivatives with respect to these parameters are computed below. 

Let us first consider the halo mass for which we get

\begin{eqnarray}
\frac{\partial \ln{P(r)}}{\log{M_{200}}} & = & \frac{\ln{10}}{3} \\ 
 & \times & \left [ 
2 - \frac{{\cal{Q}}_{x}^{GR}(r/R_{200}) + {\cal{Q}}_{x}^{MG}(r/R_{200})}
{{\cal{Q}}_{0}^{GR}(r/R_{200}) + {\cal{Q}}_{0}^{MG}(r/R_{200})} \frac{r}{R_{200}}\right ] \nonumber
\label{eq: dlnPdlogM200}
\end{eqnarray}
where ${\cal{Q}}_{0}^{GR}(x)$ and ${\cal{Q}}_{0}^{MG}(x)$ are defined in Eqs.(\ref{eq: qzgrdef}) and (\ref{eq: qzmgdef}), respectively, and ${\cal{Q}}_{x}^{GR}(x)$ and ${\cal{Q}}_{x}^{MG}(x)$ are their derivatives with respect to $x$ which we compute numerically. It is worth noting that the derivative with respect to $\log{M_{200}}$ is the only one asking for numerical differentiation, while all other cases ask for numerical integration which is more stable. For $c_{200}$, indeed, we get

\begin{equation}
\frac{\partial \ln{P(r)}}{\partial c_{200}} =     
\frac{{\cal{Q}}_{c}^{GR}(r/R_{200}) + {\cal{Q}}_{c}^{MG}(r/R_{200})}
{{\cal{Q}}_{0}^{GR}(r/R_{200}) + {\cal{Q}}_{0}^{MG}(r/R_{200})} 
\label{eq: dlnPdc200}
\end{equation}
with 

\begin{eqnarray}
{\cal{Q}}_{c}^{GR}(x) & = & \left [ \frac{c_{200}/(1 + c_{200})}
{\ln{(1 + c_{200})} - c_{200}/(1 + c_{200})} \right ]^2 \nonumber \\
 & \times & \int_{x}^{\infty}{
\frac{
{\cal{C}}(x^{\prime}, c_{200})
x^{\prime}}{(1 + c_{200} x^{\prime})^2} \tilde{n}_e(x^{\prime}) dx^{\prime}} \ ,
\label{eq: qcgrdef} 
\end{eqnarray}

\begin{equation}
{\cal{Q}}_{c}^{MG}(x) = 
\int_{x}^{\infty}{
\frac{2 c_{200} \Xi_{1} (1 - c_{200} x^{\prime})}{(1 + c_{200} x^{\prime})^4}
\tilde{n}_e(x^{\prime}) dx^{\prime}} 
\ ,
\label{eq: qcmgdef} 
\end{equation}
having defined $\tilde{n}_e(x) = n_e(x)/n_{01}$, and 

\begin{equation}
{\cal{C}}(x^{\prime}, c_{200}) = 1 - [(1 - c_{200}^{-1} (1 + c_{200})^2 \ln{(1 + c_{200})}] x^{\prime}
\end{equation}
Note that all the ${\cal{Q}}_{\mu}^{GR}(x)$ and ${\cal{Q}}_{\mu}^{MG}(x)$ defined above and in the following also depend on $c_{200}$, the double\,-\,$\beta$ profile parameters $(\beta, \log{x_{c1}}, \log{x_{c2}}, \log{(n_{02}/n_{01})})$, and $(\alpha_H. \beta_1)$ for the MG labeled quantities. We drop the dependence on them just to shorten the notation.

The derivatives with respect to the double\,-\,$\beta$ profile parameters are quite straightforward to compute, being simply given by

\begin{equation}
\frac{\partial \ln{P(r)}}{\partial \log{n_{01}}} = \ln{10} \ ,
\label{eq: dlnPdlnn01}
\end{equation}

\begin{equation}
\frac{\partial \ln{P(r)}}{\partial p_{\mu}} =     
\frac{{\cal{Q}}_{\mu}^{GR}(r/R_{200}) + {\cal{Q}}_{\mu}^{MG}(r/R_{200})}
{{\cal{Q}}_{0}^{GR}(r/R_{200}) + {\cal{Q}}_{0}^{MG}(r/R_{200})}
\label{eq: dlnPdpmu}
\end{equation}
for $p_{\mu} \in \{\beta, \log{x_{c1}}, \log{x_{c2}}, \log{f_{21}}\}$ and

\begin{equation}
\begin{array}{l}
\displaystyle{{\cal{Q}}_{\mu}^{GR}(x) =} \\ 
\\ 
\displaystyle{\int_{x}^{\infty}{
\frac{\ln{(1 + c_{200} x^{\prime}}- c_{200} x^{\prime}/(1 + c_{200} x^{\prime})}{\ln{(1 + c_{200}}- c_{200}/(1 + c_{200})}
\frac{\partial \tilde{n}_e(x^{\prime})}{\partial p_{\mu}} dx^{\prime}}} \ ,
\end{array}
\label{eq: qmugrdef}
\end{equation}

\begin{equation}
{\cal{Q}}_{\mu}^{GR}(x) = \int_{x}^{\infty}{
\frac{c_{200}^2 \Xi_1 (1 - c_{200} x^{\prime})}{(1 + c_{200} x^{\prime})^3}
\frac{\partial \tilde{n}_e(x^{\prime})}{\partial p_{\mu}} dx^{\prime}} \ .
\label{eq: qmumgdef}
\end{equation}
The derivatives of the rescaled electron density profile $\tilde{n}_e(x)$, which also enter the XR Fisher matrix, are trivial to compute so that we do not report them here.

Finally, we give below the derivatives with respect to the DHOST parameters which are

\begin{eqnarray}
\frac{\partial \ln{P(r)}}{\partial \alpha_H} & = & 
\frac{1}{1 - \alpha_H - 3 \beta_1} \\
& - &  
\frac{{\cal{Q}}_{0}^{MG}(x)}
{{\cal{Q}}_{0}^{GR}(x) + {\cal{Q}}_{0}^{MG}(x)}
\frac{\alpha_{H}^{2} + 4 \alpha_H \beta_1 + 3 \beta_{1}^{2}}
{2(\alpha_H + 2 \beta_1)^2} \ , \nonumber
\label{eq: dlnPdalphaH}
\end{eqnarray}

\begin{eqnarray}
\frac{\partial \ln{P(r)}}{\partial \beta_1} & = &   
\frac{1}{1 - \alpha_H - 3 \beta_1} \\ 
& - &  
\frac{{\cal{Q}}_{0}^{MG}(x)}
{{\cal{Q}}_{0}^{GR}(x) + {\cal{Q}}_{0}^{MG}(x)}
\frac{\beta_1 (\alpha_H + \beta_1)}
{2(\alpha_H + 2 \beta_1)^2} \ , \nonumber
\label{eq: dlnPdbeta1}
\end{eqnarray}

\begin{equation}
\frac{\partial \ln{P(r)}}{\partial \gamma_N} = 
\frac{1}{1 - \alpha_H - 3 \beta_1} \ .
\label{eq: dlnPdgammaN}
\end{equation}
Note that the derivatives with respect to $(\alpha_H, \beta_1, \gamma_N)$ are all equal to 1 for the GR fiducial.


\bibliography{bibliography} 


\end{document}